\documentstyle[12pt]{article}
\begin{document}
\titlepage
\title{\begin{flushright}
Preprint PNPI-2075,  1995
\end{flushright}
\bigskip\bigskip\bigskip
Unimodular transformations of the supermanifolds and
the calculation of the multi-loop amplitudes  in
the superstring theory}
\date{}
\author{G.S.  Danilov \thanks{E-mail address:
danilov@lnpi.spb.su}\\ Petersburg Nuclear Physics Institute,\\
Gatchina, 188350, St.-Petersburg, Russia}

\maketitle
\begin{abstract}
The modular transformations of the $(1|1)$ complex
supermanifolds in the like-Schottky modular parameterization are
discussed.  It is shown that these "supermodular"
transformations depend on the spinor structure of the $(1|1)$
complex supermanifold by terms proportional to the odd modular
parameters.  The above terms are calculated in the explicit
form. The discussed terms are important for the study of the
possible divergencies in the Ramond-Neveu-Schwarz superstring
theory. In addition, they are necessary to calculate the
dependence on the odd moduli of the fundamental domain in the
modular space.  The supermodular transformations of the
multi-loop superstring partition functions calculated by the
solution of the Ward identities are studied. In the
present paper, it is shown that the above Ward identities are
covariant under the supermodular transformations. Hence the
considered partition functions necessarily possess the
covariance under the supermodular transformations discussed. It
is demonstrated in the explicit form the covariance of the above
partition functions at zero odd moduli under those supermodular
transformations in the Ramond sector, which turn a pair of even
genus-1 spinor structures to a pair of the odd genus-1 spinor
ones. The brief consideration of the cancellation of divergences
is given.

\end{abstract}
\newpage

\section{Introduction}
In the Ramond-Neveu-Schwarz superstring theory
\cite{rnshw,gosch} the supermanifold formalism \cite{bshw}
occurs more appropriate for  multi-loop calculations than a
formalism of the Riemann surfaces.  Indeed, in the superstring
theory every $2\pi$-twist about $A$- or $B$-cycle on the Riemann
surface is, generally, accompanied by a supersymmetrical
transformation including, in addition, a boson-fermion mixing.
The above mixing can be taken into account by an extension
to the complex $(1|1)$ supermanifolds \cite{bshw} of that
complex $z$-plane where the genus-$n$ Riemann surfaces are
mapped. Every supermanifold is described by the supercoordinate
$t=(z|\theta)$ where $z$ is the complex local coordinate and
$\theta$ is its Grassmann ( odd ) partner. Grassmann ( odd )
parameters of the discussed boson-fermion mixing are expressed in
terms of complex Grassmann ( odd ) modular parameters,
which are assigned to every complex $(1|1)$
supermanifold in addition to ordinary ( even ) complex
Riemann moduli.  In this case fermion strings are
classified over "superspin" structures instead the ordinary spin
structures \cite{swit}.  The superspin structures are defined
for superfields on the discussed complex $(1|1)$ supermanifolds
\cite{bshw}.  Being twisted about $(A,B)$-cycles, the
superfields are changed by mappings that present superconformal
versions of fractional linear transformations.  Generally, every
considered mapping depends on $(3|2)$ parameters \cite{bshw}.
For odd parameters to be arbitrary, these mappings include, in
addition, fermion-boson mixing above. It differs the superspin
structures from the ordinary spin ones. Indeed, the ordinary
spin structures \cite{swit} imply that boson fields are
single-valued on Riemann surfaces. Only fermion fields being
twisted about $(A,B)$-cycles, may receive the factor of -1. For
all odd parameters to be zero, every genus-n superspin structure
$L=(l_1 ,l_2)$ is reduced to the ordinary $(l_1,l_2)$ spin one.
Here $l_1$ and $l_2$ are the theta function characteristics:
$(l_1,l_2)=\bigcup_s(l_{1s} ,l_{2s})$ where $l_{is}\in(0,1/2)$.
The (super)spin structure is even, if $4l_1l_2=4\sum_{s=1}^
nl_{1s}l_{2s}$ is even.  It is odd, if $4l_1l_2$ is odd.

One could to avoid the supermanifold formalism
\cite{bshw} using the prescription \cite{ver,martnp} for the
integration over the odd moduli. In this case,
however, multi-loop amplitudes turn out to be depended on a
choice of basis of the gravitino zero modes \cite{ver,as,momor}.
It means that the world-sheet supersymmetry is lost in the
scheme discussed.  Indeed, in the superstring theory both the
{\it vierbein} and the gravitino field are the gauge fields.
Owing to the gauge invariance the "true" superstring amplitudes
are independent of the choice of a gauge of the above gauge
fields.  Therefore, they have no dependence on the choice of
basis of the gravitino zero modes.  The discussed dependence on
the choice of basis of the gravitino zero modes appears to be a
serious difficulty in the scheme \cite{ver,as,momor}. But  the
above difficulty is absent in the supermanifold formalism
\cite{bshw} that possesses the manifest world-sheet supersymmetry.

In the considered scheme \cite{bshw} the multi-loop amplitudes
are obtained \cite{vec8,pst,ntw,dan1,dan0} by the summation over
"superspin" structure contributions integrated over both the
even moduli and the odd ones and over the vertex
supercoordinates, as well.
Every superspin structure contribution presents
the suitable partition function multiplied by the vacuum
expectation of the vertex product. The
above vacuum expectations are expressed in terms of superfield
vacuum correlators. Different approaches to the calculation of
the vacuum correlators and of the partition functions  have been
proposed \cite{vec8,pst,ntw,dan1,dan0,dan5}. In \cite{dan5}, the
superfield vacuum correlators and the partition functions have
been calculated in the explicit form for all the even superspin
structures. The integration over the modular parameters and over
the vertex supercoordinates needs an additional
investigation. Indeed, for every superspin structure
contribution, both the integral over the even moduli and the
integral over the $\{z^{(r)},\bar z^{(r)}\}$ vertex local
coordinates are divergent.\footnote{Troughout this paper the
line over denotes complex conjugation.} The divergencies of the
integrals over $\{z^{(r)},\bar z^{(r)}\}$ arise from the region
where all the vertices move to be closed together and from
the region where all they move away from each other. The
divergencies of the integrals over the modular parameters are
due to a degeneration of the Riemann surfaces. Of the main
difficulty for the investigation are the possible divergencies
due to a degeneration of genus-n Riemann surfaces $(n>1)$ into a
few ones of the lower genus. It is expected
\cite{martnp,vec8,martpl,mandel,berk} that the above
divergencies disappear after the summation over spinor
structures to be performed, but this problem needs an
additional study. In any case, the correct consideration of the
divergency problem requires even if an implicit regularization
procedure. The above regularization
procedure must be chosen ensuring the supermodular group
invariance of the superstring amplitudes. The supermodular group
does be the superconformal extension of the modular group in
the boson string theory.  Generally, the supermodular
transformations present the globally defined $t\rightarrow\hat
t(t,\{q_N\})$ holomorphic superconformal mappings \cite{bshw} of
the $t=(z|\theta)$ supercoordinate, which are accompanied by the
$q_N\rightarrow \hat q_N(\{q_N\})$ holomorphic mappings of the
complex moduli $q_N$ and, generally, by the $L\rightarrow\hat L$
change of the superspin structure, as well.  To avoid the
explicit regularization procedure, it seems attractive to write
down the multi-loop superstring amplitude in the form of the
integral over both $\{q_N,\overline q_N\}$ and $\{t^{(r)},\bar
t^{(r)}\}$ of the integrand covariant under the supermodular
transformations. Being defined by the above integral, the
considered superstring amplitude surely satisfies the
restrictions due to the supermodular group, at least, if the
above integrand has no non-integrable singularities. In this
case the discussed construction solves the problem of the
calculation of the superstring amplitudes. Simultaneously,
the study of the divergency problem is reduced to the
investigation of the singularities of the supermodular covariant
integrand.  Owing to the supermodular invariance, every
superspin structure contribution possesses covariance under the
supermodular transformations. So the desired integrand presents
the sum over all the superspin structures contributions, every
term being the partition function multiplied by the vacuum
expectation of the vertex product. The discussed scheme is,
however, complicated by the non-split in the sense of
\cite{crrab} of the supermanifolds. At least, the above
non-split takes place, if the like-Schottky modular
parameterization \cite{vec8,pst,ntw,dan1,dan0,dan5} is used. In
this case the modular group transformations ( $t\rightarrow\hat
t(t,\{q_N\})$, $q_N\rightarrow\hat q_N(\{q_N\})$,
$L\rightarrow\hat L$ ) affect not only the bodies of the modular
parameters, but on the soul components, as well.  So the
resulted modular parameters $\hat q_N(\{q_N\})$ and the resulted
supercoordinate $t(t,\{q_N\})$ depend non-trivially on the odd
modular parameters. Particular, among terms proportional to odd
modular parameters, there are terms depending on the superspin
structure $L$.  Because of the above $L$ dependence of both
$t\rightarrow\hat t(t,\{q_N\})$ and $q_N\rightarrow\hat
q_N(\{q_N\})$, the discussed integrand is non-covariant under
the supermodular group, if the $q_N$ moduli are chosen to be the
same for all the superspin structures. To build the supermodular
covariant integrand, the calculation of the $L$ dependence of
both $\hat t(t,\{q_N\})$ and $\hat q_N(\{q_N\})$ is necessary.
It seems that the knowledge of the above $L$ dependence is
necessary also, if instead of the discussed scheme, one will
attempt to construct an explicit regularization procedure for
the integration of every superspin structure contribution.

In the present paper we calculate the explicit dependence on
the odd modular parameters of both $\hat t(t,\{q_N\})$ and
$\hat q_N(\{q_N\})$. Generally, the above dependence is obtained
to be series in the odd modular parameters. We show that in both
$\hat t$ and $\hat q_N$ among terms proportional to odd modular
parameters, there are terms depending on the $L$ superspin
structure.  Furthermore, we propose method constructing the
supermodular covariant integrand in the expression for the
multi-loop amplitude.  The above integrand presents the sum over
all the superspin structures contributions, every term being
calculated at its own moduli $\{q_{NL}\}$ and its own
supercoordinates $\{t_L^{(r)}\}$, as well. These
$(q_{NL},t_L^{(r)})$ variables are functions of the
$(\{q_N\},t^{(r)})$ variables of the integration:
$q_{N_L}=(q_{N_L}(\{q_N\})$ and
$t_L^{(r)}=t_L^{(r)}(t^{(r)},\{q_N\})$.
The above $(q_{NL},t_L^{(r)})$ functions are calculated from the
condition that the same
$(t\rightarrow\hat t,q_N\rightarrow\hat q_N)$ change of
the $(t,q_N)$ variables corresponds to all the
$(t_{L}\rightarrow\hat
t_{\hat L})$ mappings associated with the  particular
supermodular transformation.  To avoid misunderstands, it is
necessary to note that the changes of  $t$ under
$2\pi$-twists about $(A_s,B_s)$-cycles remain depending on the
$L$ superspin structure. Moreover, in this case the discussed
changes of $t$ are not, generally, described by any simple
expressions similar to the Schottky transformations. It is the
fee for the supermodular transformations of $(t,q_N)$ to
be independent of $L$.  The desired supercovariant integrand
turns out to be calculated uniquely by employing only a part of
the supermodular transformations.  So, to be sure in the
self-consistency of the discussed scheme, one should verify that
the above integrand is covariant under the whole supermodular
group. This verification requires, however, an additional study
of the supermodular transformations that is planned  in  another
paper. Instead, in the present paper we discuss the changes
under the supermodular transformations of the partition
functions calculated in \cite{dan5}. For the theory to be
self-consistent, the multi-loop partition functions must be
covariant under the supermodular group. We argue that the
considered partition functions \cite{dan5} possess the
supermodular covariance required.

The discussed partition functions have been calculated
\cite{dan5} from equations \cite{dan1,dan0,dan5} that
are none other than Ward identities. These  equations fully
determine the partition functions  up to  constant factors only.
The discussed equations are derived from the condition that the
multi-loop superstring amplitudes are independent of a choice of
both the {\it vierbein} and the gravitino field.  So it seems
natural to expect that the above equations are covariant under
the supermodular group transformations. Such is indeed the case,
and in the present paper we give the direct proof of the
supermodular covariance of the equations discussed. Therefore,
the partition functions \cite{dan5} being the solution of these
equations, with necessariness satisfy restrictions due to the
supermodular group.  Unfortunately, it is difficult to obtain a
more direct evidence for the covariance of the discussed
partition functions \cite{dan5} under the whole supermodular
group. Nevertheless, one can attempt to demonstrate the
covariance  of the above partition functions  under some
subgroups of the supermodular group.  Particular, we demonstrate
that at zero odd moduli the partition functions \cite{dan5} are
covariant under the supermodular transformations, which turn a
pair of even genus-1 superspin structures in the Ramond sector
to a pair of the odd genus-1 superspin ones.

Besides the application to the divergency problem,
the dependence on the odd modular parameters of the
supermodular transformations is necessary to calculate
the dependence on the odd moduli of the region of the
integration over the even moduli in the expressions for the
multi-loop superstring amplitudes.  Indeed, the moduli being
defined modulo the supermodular group \cite{crrab}, the even
moduli are integrated over the fundamental domain that is
determined by the condition that different varieties of moduli
correspond to topologically non-equivalent supermanifolds. It is
similar to the boson string theory where the region of the
integration over moduli is determined by the modular invariance.
Inasmuch as the supermanifods are non-compact in the sense of
\cite{leites}, the boundary of the discussed fundamental domain
$\Sigma$ depends on the odd moduli.  When integrating over the
odd moduli $q_{od}$, the dependence of $\Sigma$ on odd modular
parameters must necessarily be taken into account. It is obvious
that the discussed $q_{od}$ dependence of $\Sigma$ is  just
determined by the $q_{od}$ dependence of the even moduli $\hat
q_{ev}(\{q_N\})$ obtained by the $q_N\rightarrow\hat
q_N(\{q_N\})$ supermodular transformations of
$\{q_N\}=\{q_{ev},q_{od}\}$.

The arising of the dependence on the superspin structure in both
$\hat t(t,\{q_N\})$ and $\hat q_N(\{q_N\})$ when the odd moduli
present, can be understood as it follows.  For zero odd
moduli, the supermodular transformations are reduced to the
modular ones, which form the discrete group of globally
defined holomorphic $z\rightarrow\hat z^{(0)}(z,\{q_{ev}\})$
transformations accompanied by the $q_{ev}\rightarrow \hat
q_{ev}^{(0)}(\{q_{ev}\})$ change of the $q_{ev}$ even moduli. In
this case the modular transformations
$\omega^{(r)}(\{q_{ev}\})\rightarrow \omega^{(r)}(\{\hat
q_{ev}^{(0)}\})$ of the $\omega^{(r)}(\{q_{ev}\})$ period
matrices associated with Riemann surfaces determine in an
implicit form all the new moduli $\hat q_{ev}^{(0)}$ in terms of
$q_{ev}$ up only to arbitrariness caused by possible
fractionally linear transformations of Riemann surfaces. Since
the $\omega^{(r)}$ matrix does not depend on the superspin
structure, both the $\{\hat q_{ev}\}$ sets and the $\hat
z^{(0)}(z,\{q_{ev}\})$ local coordinate appear to be the same
for all the spin structures, if the $\{q_{ev}\}$ set is chosen
to be the same for the spin structures considered. In the
presence of the odd moduli, however, period matrices are
assigned to $(1|1)$ supermanifolds rather than the Riemann
surfaces \cite{bshw,rschv}. For the genus $n\geq2$ the above
$\omega(\{q_N\};L)$ period matrices depend on the $L$ superspin
structure by terms proportional to odd moduli \cite{pst,cqg}.
These terms arise because in the considered scheme the fermions
mix the bosons under $2\pi$-twists about $(A_s,B_s)$-cycles. So
in the superstring theory, there are no reasons for $\hat
q_N(\{q_N\})$ and for $\hat t(t,\{q_N\})$ to be independent of
$L$. Moreover, though the supermodular transformations of the
above period matrices are described by the same relations as
modular transformations of period matrices in the boson string
theory, the above relations are insufficient to determine all
the resulting moduli in terms of the "old" ones, if the odd
moduli present. Only if the action of the supermodular group on
the odd moduli to be determined, the discussed relations give in
an implicit form the fundamental domain $\Sigma$  in the
modular space.

The calculation of the supermodular group action on the
odd moduli is one of goals of the present paper. In general case
the resulted $\hat q_{od}(\{q_N\})$ odd moduli are calculated in
terms of both the parameters of a suitable modular
transformation at zero odd modular parameters and the $q_{od}$
modular parameters, as well. The dependence of $\hat q_{od}$
on $q_{od}$ is obtained in the form of a series in $q_{od}$. The
$\omega(\{q_N\};L)$ period matrices in the Neveu-Schwarz sector
have been calculated in \cite{vec8,pst}. In the Ramond sector
for the even superspin structures the discussed period matrices
have been calculated in \cite{dan5}.  For the odd superspin
structures these $\omega(\{q_N\};L)$ period matrices can be
calculated by factorization of the even superspin structure in
two odd superspin structures when "handles" move away from each
other.  This calculation is planned to give in another place.

Like previous papers \cite{vec8,pst,ntw,dan1,dan0,dan5}, we use
superconformal versions of the Schottky groups
\cite{lovel,fried}. Apparently, it is the only modular
parameterization that allows to perform explicit calculations of
the partition functions in the terms of the even and odd moduli.
There are different ways to supersymmetrize ordinary spin
structures, but supersymmetrizations do not all be suitable for
the superstring theory.  Especially, because the space of
half-forms does not necessarily have a basis when there are odd
moduli \cite{hodkin}. The super-Schottky groups appropriate for
all superspin structures have been constructed in
\cite{cqg,dan3,dan4}. In the $l_{1s}=0$ case the super-Schottky
groups have been built before in \cite{martnp,vec8,pst}.  The
above $l_{1s}=0$ case corresponds to the boson loop
\cite{vec8,pst}. The boson loops can be turned into another
boson ones by the $(l_{1s}=0,l_{2s}=1/2)\rightarrow
(l_{1s}=0,l_{2s}=0)$ supermodular transformations discussed
already in \cite{cqg,dan3}. These  transformations restrict
the argument of every Schottky multiplier $k_s$, for example, as
$|\arg k_s|\leq\pi$.

Additional restrictions on the fundamental domain in the modular
space are due to the supermodular transformations that, for the
given $s$, interchange $A_s$-cycle and $B_s$-one. The above
supermodular transformations change both the moduli
$q_N$ to be $\hat q_N$ and the  $t$ supercoordinate
to be $\hat t(t,\{q_N\})$. We calculate both $\hat
t(t,\{q_N\})$ and  $\{\hat q_N\}$ in terms of the considered
transformation taken at zero odd moduli. The parameters of the
above transformation can not be calculated in the explicit form.
So we obtain the explicit dependence of $\hat t(t,\{q_N\})$ and
of $\hat q_N$ only on the odd moduli. We show that both $\hat
t(t,\{q_N\})$ and $\hat q_N$ depend on the $L$ superspin
structure. In the $l_{1s}\neq 0$ case the fermion fields are
non-periodical about the $A_s$-cycle, superfields being branched
on the complex $z$-plane where Riemann surfaces are mapped.
Hence cuts arise on the complex $z$-plane above.  Sets of these
cuts can be drown in different ways, but the varieties of these
cuts can be turn into each other by suitable supermodular
transformations.

When the cuts on the $z$-plane present,
there are the supermodular transformations due to going
$A_s$-cycles over each other. Among these transformations, of
the especial interest are those, which turn a pair
of the even genus-$1$ structures into a pair of the odd
genus-$1$ structures:  $(l_{2s_1}=0, l_{2s_2}=0)\rightarrow
(l_{2s_1}=1/2,l_{2s_2}=1/2)$, both $l_{1s_1}= l_{1s_2}=1/2$
being unchanged. If odd moduli vanish, the
moduli and the $(z|\theta)$ coordinates are
unchanged under the discussed transformations. For non-zero odd
moduli, both the moduli and the $(z|\theta)$
coordinates are changed. In this case we calculate in the
explicit form both the resulted moduli and the resulted
supercoordinates. The obtained results are
used to demonstrate the covariance of the partition functions
\cite{dan5} at zero odd moduli under the considered supermodular
transformations.

In the supersymmetrical formalism
\cite{bshw} the problem of the calculation of the partition
functions and of the superfield vacuum correlators is
concentrated, in mainly, on those superspin structures where at
least one of the $l_{1s}$ characteristics is unequal to zero.
Indeed, for superspin structures where all the $l_{1s}$
characteristics are equal to zero,  the  considered expressions
can be derived \cite{vec8} by a simple extension of the boson
string results \cite{vec7}. All the other superspin structures
can not be derived in this way.  Generally, the procedure of
"sewing" \cite{pst,ntw} allows to consider the discussed
superspin structures, but this scheme seems to be complicated,
the results being obtained in the form that is rather difficult
for an investigation. Main difficulties in the "sewing" scheme
are due to the calculation of the Ramond zero mode contributions
\cite{ntw}. The above shortcomings are absent in the scheme
developed in \cite{dan1,dan0,dan5}.  In the present paper we
show in the explicit form that at zero odd moduli, the
partition functions calculated in the considered scheme
\cite{dan1,dan0,dan5} possess supermodular covariant under
supermodular transformations turning two even genus-1 structures
to a pair of the odd genus-1 ones.

The paper is organized as it follows. In Section 2 we give the
description of the superspin structures in terms of
super-Schottky group. We discuss also the fundamental domain in
the modular space. Mainly, this Section presents a brief review
of the results \cite{cqg,dan3,dan4} essential for understanding
the following Sections. In Section 3 we consider the
supermodular transformations,  which, for the given $s$,
interchange $A_s$-cycle and $B_s$-one.  In Section 4 we consider
the supermodular transformations, which turn pair of the even
genus-$1$ spinor structures with $l_{1s_1}= l_{1s_2}=1/2$ into a
pair of the odd genus-$1$ spinor ones. The supermodular
covariance of the multi-loop partition functions is discussed in
Section 5 and in Section 6. In Section 7 the supermodular
covariant integrand in the expression for the multi-loop
amplitudes is constructed. The integration region over moduli is
defined. A brief discussion of the divergency problem is given.

\section{Superspin structures}

Generally, every  superspin structure given on a genus-n complex
$(1|1)$  supermanifold is defined by the transformations
$(\Gamma_{a,s}(l_{1s}),\Gamma_{b,s}(l_{2s}))$ that are associated
with rounds about $(A_s,B_s)$-cycles, respectively
($s=1,2,...,n$). Like the previous Section, we map the
supermanifolds by the supercoordinate $t=(z|\theta)$ where z is
a local complex coordinate and  $\theta$ is its odd partner.
Following to \cite{martnp,vec8,pst,dan1,dan0} we use for
$\Gamma_{b,s}(l_{2s})$ the superconformal versions of Schottky
transformations. For zero odd modular parameters
these transformations $\Gamma_{b,s}^{(o)}(l_{2s})$ are defined
as
\begin{equation}
\Gamma_{b,s}^{(o)}(l_{2s})=\left\{z\rightarrow   g_s(z),\quad
\theta\rightarrow
-\frac{(-1)^{l_{2s}}\theta}{c_sz+d_s}\right\}
\label{schot}
\end{equation}
where $g_s(z)$ is the Schottky transformations:
\begin{equation}
g_s(z)=\frac{a_sz+b_s}{c_sz+d_s} \quad {\rm{with}}
\quad a_sd_ s-b_sc_s=1.
\label{sch}
\end{equation}
Eq. (\ref{schot}) takes into account \cite{martnp} that for
$l_{2s}=0$, the spinors are multiplied by -1. Furthermore,
the $( a_s ,b_s ,c_s ,d_s)$ parameters can be expressed
\cite{fried} in  terms of two fixed points
$u_s$ and $v_s$ on the complex $z$-plane  together with  the
multiplier $k_s$ by
\begin{equation}
a=\frac{u-kv}{\sqrt k(u-v)},\quad d=\frac{ku-v}
{\sqrt k(u-v)}\quad{\rm and}\quad c=\frac{1-k}{\sqrt k(u-v)}
\label{uvk}
\end{equation}
(index $s$ is omitted).
Every transformation (\ref{schot})
turns the circle $C_{v_s}$  into
$C_{u_s}$ where
\begin{equation}
C_{v_s}=\{z:|c_sz+d_s|=1\}\quad{\rm and}\quad
C_{u_s}=\{z:|-c_sz+a_s|=1\}.
\label{circ}
\end{equation}
It is useful to note that
every $v_s$ point is situated inside of $C_{v_s}$ circle
and every $u_s$ point is situated inside of $C_{u_s}$.
The exterior of all the circles above is chosen to be the
fundamental domain $\Omega$ on the complex $z$-plane. A path
about $C_{v_s}$- circle ( or about $C_{u_s}$-circle )
corresponds to $2\pi$-twist about $A_s$-cycle.  Under the above
path the spinors are multiplied by -1 in the $l_{1s}=1/2$ case,
for $l_{1s}=0$ they being unchanged \cite{martnp}.  Therefore,
$2\pi$-twists about $A_s$-cycles are associated with the
following $\Gamma_{a,s}^{(o)}(l_{1s})$ mappings:
\begin{equation}
\Gamma_{a,s}^{(o)}(l_{1s})=\{z\rightarrow z,\quad
\theta\rightarrow(-1)^{2l_{1s}}\theta\}.
\label{around}
\end{equation}
To extend  the discussed mappings (\ref{schot}) and
(\ref{around}) to arbitrary odd moduli it is necessary to find a
relation between odd parameters in $\Gamma_{a,s}(l_{1s}=1/2)$
and those in $\Gamma_{b,s}(l_{2s})$. Especially, because in
the general case  the space of half-forms does not have a basis
when there are odd moduli \cite{hodkin}. To derive the desired
relation, we employ \cite{dan5,cqg,dan3,dan4} that for genus
$n=1$, there are no odd moduli.  Indeed, the genus-1 amplitudes
are obtained in terms of ordinary spin structures
\cite{swit}. Hence for every particular $s$, all the odd
parameters in both $\Gamma_{a,s}(l_{1s})$ and
$\Gamma_{b,s}(l_{2s})$ can be reduced to zero by a suitable
transformation $\tilde\Gamma_s$, which is the same for both the
transformations above:
\begin{equation}
\Gamma_{a,s}(l_{1s})=
\tilde\Gamma_s^{-1}\Gamma_{ a ,s}^{(o)}(l_{1s})
\tilde\Gamma_s, \qquad \Gamma_{b,s}(l_{2s})=
\tilde\Gamma_s^{-1}\Gamma_{b ,s}^{(o)}(l_{2s})\tilde\Gamma_s
\label{super}
\end{equation}
where $\Gamma_{a,s}^{(o)}(l_{1s})$ is given by (\ref{around}),
$\Gamma_{b,s}^{(o)}(l_{2s})$ is given by (\ref{schot}) and
$\tilde\Gamma_s$ depends, in addition, on two odd
parameters, they being $(\mu_s,\nu_s)$.  We choose
\cite{dan5,cqg,dan3,dan4} the $\tilde\Gamma_s$ mapping as
\begin{eqnarray}
\tilde\Gamma_s:\qquad
z\rightarrow z_s+\theta_s\varepsilon_s(z_s),\qquad
\theta\rightarrow\theta_s(1+\varepsilon_s
\varepsilon_ s'/2)+
\varepsilon_s(z_s);\nonumber\\
\varepsilon_ s'=\partial_{z}\varepsilon_s(z),\qquad
\varepsilon_s(z)=[\mu_s( z-v_s)-
\nu_s( z-u_s](u_s-v_s)^{-1}.
\label{gammat}
\end{eqnarray}
In this case the $\Gamma_{b,s}(l_{2s}=1/2)$
mappings appear to be identical to those proposed in
\cite{martnp,vec8}. In the explicit form the discussed
$\Gamma_{b,s}(l_{2s}=1/2)$ mappings are given in
\cite{martnp,vec8,pst,dan0,dan5,cqg}. They can also be written
down as
\begin{eqnarray}
\Gamma_{b,s}(l_{2s}):\quad
z\rightarrow g_s(z)+\frac{\theta\epsilon_s(z,l_{2s})}
{(c_sz+d_s)^2}
-\frac{(-1)^{2l_{2s}}
\varepsilon_s(z)\varepsilon_s'(z)[z-g_s(z)]}
{(c_sz+d_s)},\nonumber\\
\theta\rightarrow-
\frac{(-1)^{2l_{2s}}\theta[1+\varepsilon_s(z)\varepsilon_s'(z)]+
\epsilon_s(z,l_{2s})}{c_sz+d_s}-
\frac{\theta\varepsilon_s(z)\varepsilon_s'(z)}
{(c_sz+d_s)^2}
\label{mart}
\end{eqnarray}
where $g_s(z)$ is the Schottky transformation (\ref{sch}).  Both
$\varepsilon_s(z)$ and $\varepsilon_s'(z)$ are
defined in (\ref{gammat}). In addition, $\epsilon_s(z,l_{2s})$
is defined by
\begin{equation}
\epsilon_s(z,l_s)=-(-1)^{2l_s}
(c_sz+d_s)\varepsilon_s\left(g_s(z)\right)-
\varepsilon_s(z).
\label{eps}
\end{equation}
Eq.(\ref{mart}) shows that $\Gamma_{b,s}(l_{2s}=0)$
are obtained from $\Gamma_{b,s}(l_{2s}=1/2)$ by the $\sqrt
k_s\rightarrow-\sqrt k_s$ replacement
\cite{dan5,cqg,dan3,dan4}. Employing (\ref{gammat}), one can
prove that transformations (\ref{super}) remain to be fixed the
supermanifold points $(u_s|\mu_s)$ and $(v_s|\nu_s)$, and
that $k_s$ is the multiplier of the $\Gamma_{b,s}(l_{2s}=0)$
transformation. Furthermore, it is obvious from (\ref{around})
and (\ref{super}) that $\Gamma_{a,s}^2(l_{1s})$ is the identical
transformation, as well as $\Gamma_{a,s}(l_{1s}=0)$:
$\Gamma_{a,s}^2(l_{1s})=I$, $\Gamma_{a,s}(l_{1s}=0)=I$.
Simultaneously, it is follows from (\ref{around})-(\ref{gammat})
that $\Gamma_{a,s}(l_{1s}=1/2)$ is given by
\begin{equation}
\Gamma_{a,s}(l_{1s}=1/2)= \{z\rightarrow
z-2\theta\varepsilon_s(z),
\quad\theta\rightarrow-\theta(1+2\varepsilon_s
\varepsilon_s')+2\varepsilon_s(z)\}.
\label{gammaa}
\end{equation}
It is useful to note that the right side of (\ref{gammaa}) is
equal to $\Gamma_{b,s}(l_{2s}=1/2)$ at
$\sqrt{k_s}=-1$.  Since $\Gamma_{a,s}(l_{1s}=1/2)\neq I$ and
$\Gamma_{a,s}^2(l_{1s}=1/2)=I$, a square root cut on the
considered z-plane is associated with every $l_{1s}\neq0$.
One of its  endcut points is placed inside the $C_{v_s}$
circle and another endcut point is placed inside the $\hat
C_{u_s}$ one. Explicit formulae for conformal tensors
\cite{martnp,fried,vec7} show that the above endcut points are
situated at $v_s$ and $u_s$, respectively.

Superconformal $p$-tensors $T_p(t)$ being considered, every
$\Gamma_{a,s}(l_{1s}=1/2)$ transformation relates $T_p(t)$ with
its value $T_p^{(s)}(t)$ obtained from $T_p(t)$ by $2\pi$-twist
about $C_{v_s}$-circle (\ref{circ}). So, $T_p(t)$ is
changed under the $\Gamma_{a,s}(l_{1s})= \{t\rightarrow t_s^a\}$
and $\Gamma_{b,s}=\{t\rightarrow t_s^b\}$ mappings as
\begin{equation}
T_p(t_s^a)=T_p^{(s)}(t)Q_{\Gamma_{a,s}}^p(t),\qquad
T_p(t_s^b)=T_p(t)Q_{\Gamma_{b,s}}^p(t).
\label{stens}
\end{equation}
where $Q_{\Gamma_{b ,s}}(t)$ and $Q_{\Gamma_{a ,s}}(t)$ are the
factors, which the spinor derivative $D(t)$ receives under the
$\Gamma_{b,s}l_{2s}$ mapping, and, respectively, under the
$Q_{\Gamma_{a ,s}}(t)$ one. The $D(t)$ spinor derivative is
defined as
\begin{equation}
D(t)=\theta\partial_z+\partial_\theta.
\label{supder}
\end{equation}
In (\ref{supder}) the $\partial_\theta$ derivative is meant to be
the "left" one.
For an arbitrary superconformal mapping $\Gamma=\{t\rightarrow
t_\Gamma=(z_\Gamma(t)|\theta_ \Gamma(t))\}$, the $Q_\Gamma(t)$
factor is given by
\begin{equation}
Q_\Gamma^{-1}(t)=D(t)\theta_\Gamma(t)\quad;\quad
D(t_\Gamma)= Q_\Gamma(t)D(t).
\label{supfac}
\end{equation}
It is obvious from (\ref{super}) that all the even genus-1
superspin structures can be derived by supermodular
transformations of a fixed structure because it can be done if
the odd parameters are zero. Moreover, in this case the
half-forms, as well as all vacuum superfield correlators
associated with every superspin structure, can be derived by
transformations (\ref{super}) from those taken at zero odd
parameters. Hence, at least for $n=1$, there is no
the problem of constructing a basis of the half-forms, which has
been observed in \cite{hodkin}. The discussed half-forms can be
constructed \cite{dan5} also for the higher genus
supermanifolds. It is fairly clear because the above
supermanifolds are all degenerated, in essential part, to the
genus-1 supermanifolds when handles move far from each other.
In addition, all even ( odd ) superspin structures can be
derived by supermodular transformations of a fixed even ( odd )
structure that is the necessary condition of the supermodular
invariance of the multi-loop amplitudes.

At it is was noted in the Introduction, the $t$ supercoordinate
is transformed under the supermodular group by  holomorphic
supersymmetrical transformations
$t\rightarrow\hat t(t;\{q_M\})$. Simultaneously, $q_N\rightarrow
\hat q_N(\{q_M\})$. Also, generally, the above transformations
turn out the superspin structures into each other: $L\rightarrow
\tilde L$. In the theory of Riemann surfaces  the action of the
modular group on the modular parameters can be given in an
implicit form by the relations between the
$\omega^{(r)}(\{q_{ev}\})$ period matrices and those obtained by
the action on $\omega^{(r)}(\{q_{ev}\})$ of the modular group.
The above relations are as follows
\begin{equation}
\omega^{(r)}(\{q_{ev}\})=[A\omega^{(r)}(\{\hat q_{ev}^{(0)}\})+B]
[C\omega^{(r)}(\{\hat q_{ev}^{(0)}\})+D]^{-1}
\label{modtr}
\end{equation}
where $A,B,C$ and $D$ are integral matrices \cite{martpl}
( see also \cite{dan0} ). The $\omega^{(r)}$ period matrices
in terms of the Schottky group parameters have been calculated in
\cite{fried,vec7}. The above matrices are given in the
Appendix B of the present paper.  In the genus $n>3$ case a
number of equations (\ref{modtr}) being $n(n+1)/2$, is greater
than a number $3n-3$ of the complex moduli, but only $3n-3$
among the equations are independent of each other. So eqs.
(\ref{modtr}) determine in an implicit form all the new  $\hat
q_{ev}^{(0)}$ Schottky parameters in  terms of the old ones
$\{q_{ev}\}$ up to arbitrariness due to possible fractionally
linear transformations of Riemann surfaces. To determine in the
similar way the action on the even super-Schottky group
parameters of the supermodular group, one must add (\ref{modtr})
by the calculation of a dependence of $\hat q_{ev}$ on odd
modular parameters. As it explained in two following Sections,
the above dependence is determined simultaneously with the
calculation of the action of the supermodular group on the odd
super-Schottky group parameters. One can also use, instead of
(\ref{modtr}), supermodular transformations of the period
matrices assigned to complex $(1|1)$ supermanifolds.
Supermodular transformations
$\omega(\{q_N\};L)\rightarrow\omega(\{\tilde q_N\};\tilde L)$ of
the above matrices have the same form (\ref{modtr}) as in the
theory of the Riemann surfaces:
\begin{equation}
\omega(\{q_N\};L)=[A\omega(\{\hat q_N\};\hat
L)+B][C\omega(\{\hat q_N\};\hat L)+D]^{-1}\,.
\label{smodtr}
\end{equation}
In the super-Schottky  parameterization the
above $\omega(\{q_N\};L)$ matrices in the Neveu-Schwarz sector
have been calculated  in \cite{vec8,pst}.  In the Ramond sector
the discussed period matrices for the even superspin structures
have been calculated in \cite{dan5}. The period matrices for the
odd superspin structures  can be calculated in the similar way.
This calculation is planned in another paper. For all the even
superspin structures the above period matrices is presented in
Appendix B.

To determine even if in the implicit form the action of the
supermodular group on the even super-Schottky group parameters,
eqs. (\ref{smodtr}) must be added by expressions of the odd
super-Schottky parameters $\hat q_{od}$ in terms of $\{q_N\}$.
The simplest supermodular transformations are those, which turn
the $(l_{1s}=0,l_{2s}=1/2)$ characteristics to
$(l_{1s}=0,l_{2s}=0)$ and conversely. The above transformations
have already been discussed in \cite{cqg,dan3}.  Under the
considered transformations, $\arg k_s$ is replaced by $\arg
k_s+2\pi$, other modular parameters being unchanged, as well as
the $t$ supercoordinate. Indeed, as it has been explained above,
the $\Gamma_{b,s}(l_{2s}=0)$ transformations are obtained from
the $\Gamma_{b,s}(l_{2s}=1/2)$ transformations (\ref{mart}) just
by the  $\arg k_s\rightarrow\arg k_s+2\pi$ replacement.  Hence
the considered transformations restrict the argument of every
$k_s$, for example, as $|\arg k_s|\leq\pi$.
Additional restrictions on the fundamental domain in the modular
space arise from supermodular transformations discussed in
Section 3 and Section 4. Unlike the above considered
transformations, these transformations affect the super-Schottky
group parameters. In addition, they appear depending on the
superspin structure.

\section{Interchanging $A_s$-cycles and $B_s$-cycles.}

In this Section we consider those supermodular transformations,
which, for a number of the handles, interchange $A$-cycle and
$B$-one. These transformations are associated
with the $t\rightarrow \hat t(t,\{q_N\})$ supersymmetrical
mappings of the $t=(z,\theta)$ supercoordinate as follows
\begin{equation}
\hat
z=f(z)+f'(z)\theta\xi(z)\quad{\rm and}\quad
\hat\theta=\sqrt{f'(z)}\left[\left(1+
\frac{1}{2}\xi(z)\xi'(z)\right)
\theta+\xi(z)\right]
\label{smtr}
\end{equation}
where $f(z)$ ( respectively, $\xi(z)$ ) is an ordinary
(respectively, Grassmann ) holomorphic function
\cite{bshw,rschv}. Below it is implied that $z$ in (\ref{smtr})
belongs to the fundamental domain $\Omega$, which is the
exterior of all the circles (\ref{circ}). We calculate the
dependence of both $f(z)$ and $\xi(z)$ on the odd modular
parameters for the supermodular transformations in question.
It is obvious that the
discussed transformations are determined up to the
superconformal fractionally linear
transformations. We fix the solution of (\ref{smtr}) by
the condition that the above transformations
remain unchanged the $\{N_0\}$ set of $(3|2)$  of the
super-Schottky parameters chosen to be no moduli, which is
the same for all the genus-n supermanifolds.

For the resulted
superspin structure, $2\pi$ twists
about $(A_s,B_s)$-cycles are associated with the
$(\hat\Gamma_{a,s}(\hat l_{1s}),\hat\Gamma_{b,s}(\hat l_{2s}))$
transformations instead of
$(\Gamma_{a,s}(l_{1s}),\Gamma_{b,s}(l_{2s}))$.  The above
transformations are defined by eqs.(\ref{super})-(\ref{gammaa})
in terms of the resulted $(\hat k_s,\hat u_s,\hat
v_s,\hat\mu_s,\hat\nu_s)$ Schottky parameters instead of the
$(k_s,u_s,v_s,\mu_s,\nu_s)$ Schottky ones. In this case both
$g_s(z)$, $\varepsilon(z)$ and $\epsilon(z,l_s)$ are replaced by
$\hat g_s(\hat z)$, $\hat\varepsilon(\hat z)$ and
$\hat\epsilon(z,l_s)$, respectively. Above $\hat g_s(\hat z)$,
$\hat\varepsilon(\hat z)$ and $\hat\epsilon(z,l_s)$ are
expressed in terms of the resulted moduli just the same as
$g_s(z)$, $\varepsilon(z)$ and $\epsilon(z,l_s)$ are
expressed in terms of $(k_s,u_s,v_s,\mu_s,\nu_s)$. As it was
already explained in the Introduction, we calculate both  $\hat
t$ and the $\{\hat k_s,\hat u_s,\hat v_s,\hat\mu_s,\hat\nu_s\}$
in term of $\hat z^{(0)}$ and of $\{\hat k_s^{(0)},\hat
u_s^{(0)},\hat v_s^{(0)}\}$, which are equal to above $\hat z$
and $\{\hat k_,\hat u_s,\hat v_s\}$ taken at zero odd modular
parameters.

Without loss of generality, we can assume that under the
transformations considered, $(l_{1s},l_{2s})$-characteristics
are changed only for $s=1,2...p$, they being unchanged for
$p<s\leq n$. Hence in the $s\leq p$ case every $2\pi$-twist
about $A_s$-cycle is turned to $2\pi$-twist about $B_s$-cycle
and conversely. And every $2\pi$-twist about $A_s$-cycle (
$B_s$-cycle ) is turned to itself in the $s>p$ case. Hence the
set of the equations arises as follows
\begin{eqnarray}
\hat\Gamma_{a,s}(l_{2s})(\hat t)=\hat
t\left(\Gamma_{b,s}(l_{2s})(t)\right)\quad {\rm and}\quad
\hat\Gamma_{b,s}(l_{1s})(\hat t)=\hat t^{(s)}
(\left(\Gamma_{a,s}(l_{1s})(t)\right)) \quad {\rm for}\quad
s\leq p \nonumber\\
\hat\Gamma_{b,s}(l_{2s})(\hat t)= \hat
t\left(\Gamma_{b,s}(l_{2s})(t)\right)\quad {\rm and}\quad
\hat\Gamma_{a,s}(l_{1s})(\hat t)= \hat
t^{(s)}\left(\Gamma_{a,s}(l_{1s})(t)\right)\quad {\rm for}\quad
s>p.
\label{main}
\end{eqnarray}
The $\hat t^{(s)}(t)$ value in (\ref{main}) obtained by
$2\pi$-twist of $\hat t(t)$ about $C_{v_s}$-circle (\ref{circ})
on the complex $z$-plane. We write down
(\ref{main}) in term of $f_0(z)$ and $\hat g_s^{(0)}(f)$
presenting $f(z)$ and, respectively, $\hat g_s(f)$ calculated at
zero modular parameters. These $f_0(z)$ and $\hat g_s^{(0)}(f)$
obey eqs.(\ref{main}) taken at zero odd parameters. The above
equations can be written down as follows
\begin{eqnarray}
f_0(z)-f_0(g_s)=\hat g_s^{(0)}(f_0)-f_0^{(s)}(z)=0
\quad{\rm for}\quad s\leq p, \nonumber \\
\hat g_s^{(0)}(f_0)-f_0(g_s)=f_0-f_0^{(s)}=0 \quad{\rm for}\quad
s>p
\label{mainzer}
\end{eqnarray}
where $g_s\equiv g_s(z)$ and $f_0\equiv f_0(z)$.
Furthermore, $f_0^{(s)}(z)$ is derived from $f_0(z)$ by
$2\pi$-twist about $A_s$-cycle.  Eqs.(\ref{mainzer}) determine
both $f_0(z)$ and $\hat g_s^{(0)}(f)$ up to fractionally linear
transformations.  We fix the choice of both $f_0(z)$ and $\hat
g_s^{(0)}(f)$ by the condition that the above $\{N_0\}$ set of
the Schottky parameters is unchanged. In this case both $f(z)$
and $\hat g_s(f)$ in question differ from $f_0(z)$ and $\hat
g_s^{(0)}(f)$ only by terms proportional to odd modular
parameters.  It will be convenient to define $y(z)$ and $h_s(f)$
functions as follows
\begin{equation}
y(z)=\frac{f(z)-f_0(z)}{f_0'(z)}\quad{\rm
and}\quad h_s(f)=\hat g_s(f)-\hat g_s^{(0)}(f).
\label{yz}
\end{equation}
Every of eqs.(\ref{main}) presents the set of two equations,
every equation being the  first order polynomial in $\theta$.
So, there are four equations associated with every $2\pi$-twist,
but one can verify that two of these equations follow from two
other ones.  The full set of the independent equations can be
chosen as follows. Firstly, we use the relations, which
determine $\hat z$ at $\theta=0$.  Eq.(\ref{mainzer}) being
taken into account, the above relations for $\hat z$ at
$\theta=0$ can be written down as
\begin{eqnarray}
y(z)-y(g_s)(c_sz+d_s)^2=-\frac{h_s(f)}{f_0'(g_s)g_s'(z)}+
\rho_s^{(bb)}(z) \quad{\rm for}\quad
s>p \nonumber \\
y(z)-y(g_s)(c_sz+d_s)^2=
\rho_s^{(ab)}(z) \quad{\rm for}\quad s\leq p,\nonumber\\
y(z)-y^{(s)}(z)=-\frac{h_s(f)}{{f_0^{(s)}}'(z)}+
\rho_s^{(ba)}(z) \quad{\rm
for}\quad s\leq p \nonumber\\
y(z)-y^{(s)}(z)=\rho_s^{(aa)}(z) \quad{\rm for}\quad
s>p
\label{main1}
\end{eqnarray}
where $f\equiv f(z)$ and $h_s(f)$ is defined by (\ref{yz}).
Every $y^{(s)}(z)$ in (\ref{main1}) is obtained by $2\pi$-twist
of $y(z)$ about $A_s$-cycle. The explicit formulae for the
$\rho^{(pq)}$ functions ( with $p=a,b$ and $q=a,b$ ) are given
in Appendix A. Eqs.(\ref{main1}) must be
complected by the equations, which follow from  the relations
(\ref{main}) for $\hat\theta$. To derive these equations, we
calculate $\hat g_s'(f)=\partial_f\hat g_s(f)$ using for this
purpose the linear in $\theta$ terms in the relations discussed.
We substitute this $\hat g_s'(f)$ to the relations
determining $\hat\theta$ at $\theta=0$. In this case the desired
equations turn out to be as follows
\begin{eqnarray}
\xi(z)+(-1)^{2l_{2s}}\xi(g_s)(c_sz+d_s)=\epsilon_s(z,l_{2s})+
\frac{[1-(-1)^{2l_{2s}}]\hat\varepsilon_s(f)}{\sqrt{f'(z)}}
+\eta_s^{(ab)}(z) \quad{\rm for}\quad s\leq p, \nonumber \\
\xi(z)+(-1)^{2l_{2s}}\xi(g_s)(c_sz+d_s)=\epsilon_s(z,l_{2s})
-\frac{\hat\epsilon_s(f,l_{2s})}{\sqrt{f'(z)}}
+\eta_s^{(bb)}(z) \quad{\rm for}\quad s>p  \nonumber\\
\xi(z)-(-1)^{2l_{1s}}\xi^{(s)}(z)=-[1-(-1)^{2l_{1s}}]
\varepsilon_s(z)
-\frac{\hat\epsilon_s(f,l_{1s})}{\sqrt{f'(z)}}
+\eta_s^{(ba)}(z)
\quad{\rm for}\quad s\leq p \nonumber\\
\xi(z)-(-1)^{2l_{1s}}\xi^{(s)}(z)= -[1-(-1)^{2l_{1s}}]
\left[\varepsilon_s(z)-
\frac{\hat\varepsilon_s(f)}{\sqrt{f'(z)}}\right]
+\eta_s^{(aa)}(z)
\quad{\rm for}\quad s>p
\label{main2}
\end{eqnarray}
where $f\equiv f(z)$ and $g_s\equiv g_s(z)$. The $\eta_s^{(pq)}$
functions with $p=a,b$ and $q=a,b$ are given in Appendix A.
Every $\xi^{(s)}(z)$ value in (\ref{main2}) is obtained by
$2\pi$-twist of $\xi(z)$ about $C_{v_s}$-circle
(\ref{circ}) on the complex $z$-plane. The first pair of the
equations in (\ref{main1}) and in (\ref{main2}) is derived from
those of eqs.(\ref{main}), which associated with $2\pi$-twists
about $B_s$-cycles on $z$-plane.  And the second pair of the
discussed equations in (\ref{main1}) and in (\ref{main2}) is
derived from those of eqs.(\ref{main}), which associated with
$2\pi$-twists about $A_s$-cycles on $z$-plane.  In deriving
(\ref{main1}) and (\ref{main2}), eqs.(\ref{mart}) and
(\ref{gammaa}) are used. It is follows from
(\ref{main1}) and from (\ref{main2}) that in the $l_{1s}\neq0$
case both $\xi(z)$ and $y(z)$ being twisted about $C_{v_s}$ and
$ C_{u_s}$-circles (\ref{circ}), are branched on the complex
$z$-plane. One of endcut points is placed inside the $C_{v_s}$
circle and the other endcut point is placed inside the $C_{u_s}$
one. If $f_0(z)$ is assumed to be known, eqs.(\ref{main1}), and
(\ref{main2}) can be used to calculate both $\xi(z)$ and $y(z)$.
To solve the above equations we transform them to the
integral equations.  For this purpose we build the holomorphic
Green functions $G_{gh}^{(b)}(z,z')$ and $G_{gh}^{(f)}(z,z')$,
which are conformal 2-tensors and, respectively, a conformal
3/2-tensors under $z'\rightarrow g_s(z')$ transformations on
$z'$-plane. We require also that the above Green functions have
no singularities in the $\Omega$ fundamental domain on both
$z$-plane and $z'$-one, except only at
$z=z'$.  Particular, being a non-singular conformal 2-tensor,
$G_{gh}^{(b)}(z,z')$  decreases not slower than $(z')^{-4}$ at
$z'\rightarrow\infty$. And $G_{gh}^{(f)}(z,z')$ being a
non-singular conformal 3/2-tensor, decreases not slower than
$(z')^{-3}$ at $z'\rightarrow\infty$.  It is useful to note that
under $z\rightarrow g_s(z)$ transformations the above Green
functions necessarily have the depending on $z$ periods. So they
are not to be conformal tensors  under the above
transformations.  The discussed Green functions can be obtained
from ghost Green functions considered in \cite{dan1,dan0,dan5}.
These $G_{gh}(t,t')$ Green functions appear in the special ghost
scheme \cite{dan1,dan0} that allows to calculate, among of other
things, both the moduli volume form and zero mode contributions
to the partition functions. Both $G_{gh}^{(b)}(z,z')$ and
$G_{gh}^{(f)}(z,z')$ are calculated from the $G_{gh}(t,t')$
defined in \cite{dan5} by taking all modular parameters to be
zero.  In this case, $G_{gh}(t,t')$ can be written down in terms
of $G_{gh}^{(b)}(z,z')$ and $G_{gh}^{(f)}(z,z')$ as
$G_{gh}(t,t')=G_{gh}^{(b)}(z,z')\theta'+ \theta
G_{gh}^{(f)}(z,z')$.
We normalize both $G_{gh}^{(b)}(z,z')$ and $G_{gh}^{(f)}(z,z')$
as follows
\begin{equation}
G_{gh}^{(b)}(z,z')\rightarrow-(z-z')^{-1}\quad{\rm and}\quad
G_{gh}^{(f)}(z,z')\rightarrow (z-z')^{-1}
\quad{\rm at}\quad z\rightarrow z'.
\label{norm}
\end{equation}
Then the changes of the considered Green functions under the
$z\rightarrow g_s(z)$ transformation are given by \cite{dan5}
\begin{eqnarray}
G_{gh}^{(f)}(g_s(z),z')=(-1)^{2l_{2s}-1}(c_sz+d_s)^{-1}
\left(G_{gh}^{(f)}(z,z')+
{\sum_{F_s}}' P_{F_s}(z)\tilde\chi_{F_s}^{(0)}(z')\right)
\nonumber \\
G_{gh}^{(b)}(g_s(z),z')=(c_sz+d_s)^{-2}\left(G_{gh}^{(b)}(z,z')+
{\sum_{R_s}}'P_{R_s}(z)\tilde\chi_{R_s}^{(0)}(z')\right),
\label{chgr}
\end{eqnarray}
where $\tilde\chi_{R_s}^{(0)}(z')$ are the conformal 2-tensor
zero modes ( in number of $3n-3$ ), and
$\tilde\chi_{F_s}^{(0)}(z')$ are the conformal 3/2-tensor zero
modes ( in number of $2n-2$ ).\footnote{ In terms of the
$\tilde\chi_{N_s}^{(0)}(t')$ zero modes defined in \cite{dan5},
the above $\tilde\chi_{R_s}^{(0)}(z')$ and
$\tilde\chi_{F_s}^{(0)}(z')$ are expressed as
$\theta'\tilde\chi_{R_s}^{(0)}(z')= \tilde\chi_{R_s}(t')$ and
$\tilde\chi_{F_s}^{(0)}(z')=\tilde\chi_{F_s}(t')$,
$\tilde\chi_{N_s}(t')$ being taken at zero odd modular
parameters.} We use for both $R_s$ and $F_s$ the same
notation as for the Schottky parameters:  $R_s=(k_s,u_s,v_s)$
and $F_s= (\mu_s,\nu_s)$. The summation in (\ref{chgr}) is
performed over only those $R_s=(k_s,u_s,v_s)$ and
$F_s=(\mu_s,\nu_s)$, which do not belong to the $\{N_0\}$ set
chosen to be the same for all the genus-$n$ surfaces.
Furthermore, $P_{F_s}(z)$ and $P_{R_s}(z)$ in (\ref{chgr})
present polynomials of degree-1 and, respectively, of degree-2.
In the explicit form\footnote{ In terms of the
$Y_{b,N_s}^{(0)}(t)$ polynomials given in \cite{dan5}, the above
polynomials are expressed as $P_{R_s}(z)=Y_{b,N_s}^{(0)}$ and
$\theta P_{F_s}(z)=Y_{b,F_s}^{(0)}$} the above polynomials are
given by \cite{dan1,dan0,dan5,dan9}
\begin{eqnarray}
P_{k_s}=(c_sz+d_s)^2\frac{\partial g_s(z)}{\partial
k_s},\quad P_{u_s}=(c_sz+d_s)^2\frac{\partial
g_s(z)}{\partial u_s},
\quad P_{v_s}=
(c_sz+d_s)^2\frac{\partial g_s(z)} {\partial v_s},\nonumber \\
P_{\mu_s}(z)=\frac{\partial\epsilon_s(z,l_{2s})}
{\partial\mu_s}=\frac{-2(-1)^{2l_{2s}}
[1+(-1)^{2l_{2s}}\sqrt{k_s}](z-v_s)}
{\sqrt{k_s}(u_s-v_s)}
\quad{\rm and}\quad
P_{\nu_s}(z)
\nonumber \\
=\frac{\partial\epsilon_s(z,l_{2s})}
{\partial\nu_s}
=\frac{2[1+(-1)^{2l_{2s}}\sqrt{k_s}](z-u_s)}
{(u_s-v_s)}
\label{poly}
\end{eqnarray}
where $P_{k_s}=P_{k_s}(z)$,
$P_{u_s}=P_{u_s}(z)$, $P_{v_s}=P_{v_s}(z)$, and
$\epsilon_s(z,l_{2s})$ is defined by (\ref{eps}). It is obvious
that the $G_{gh}^{(f)}$ Green functions depend on the spin
structure. In the $l_{1s}\neq0$
case they change \cite{dan5} also under $2\pi $-twist about
$C_{v_s}$ cycle as follows
\begin{equation}
G_{gh}^{(f)}(z,z')=-\frac{[1-(-1)^{2l_{1_s}}]}{2}
\left[G_{gh}^{(f)(s)}(z,z')+
{\sum_{F_s}}' P_{F_s}^{(a)}(z)\tilde\chi_{F_s}(z')\right]
\label{rgr}
\end{equation}
where $G_{gh}^{(f)(s)}(z,z')$ denotes the $G_{gh}^{(f)}$
Green function $2\pi$-twisted about $A_s$-cycle. Like
(\ref{chgr}), the summation in (\ref{rgr}) is performed over
only those $F_s=(\mu_s,\nu_s)$, which chosen to be moduli, and
$\tilde\chi_{F_s}(z')$ are the same as in (\ref{chgr}).
Furthermore, $P_{F_s}^{(a)}(z)$ are degree-1 polynomials in $z$.
The above polynomials are equal to $P_{F_s}(z)$
polynomials in eq. (\ref{poly}) taken at
$(-1)^{2l_{2s}}\sqrt{k_s}=-1$. In the explicit form
\begin{equation}
P_{\mu_s}^{(a)}(z)=-\frac{4(z-v_s)}{u_s-v_s}\quad{\rm and}\quad
P_{\nu_s}^{(a)}(z)=\frac{4(z-u_s)}{u_s-v_s}.
\label{pola}
\end{equation}
It is proved in \cite{dan5} that eqs.(\ref{chgr}) and
eqs.(\ref{rgr}) are self-consistent. In \cite{dan5} the
discussed Green functions were considered only for even spin
structures, but they can be extended without any difficulties to
odd spin ones of genus $n>1$.

To derive the set of integral equations for both $\xi(z)$
and $y(z)$ in question we start with the following relations
\begin{equation}
\xi(z)=-\int_{C(z)}G_{gh}^{(f)}(z,z')\xi(z')\frac{dz'}{2\pi i}
\quad{\rm and}\quad
y(z)=\int_{C(z)}G_{gh}^{(b)}(z,z')y(z')\frac{dz'}{2\pi i}
\label{inf}
\end{equation}
where infinitesimal contour $C(z)$ gets around $z$-point in the
positive direction. Then we deform this contour to those, which
surround both $C_{v_s}$ and $C_{u_s}$
circles (\ref{circ}) together with the $\tilde C_s$ cuts
that, generally, present because both $y(z)$ and $\xi(z)$ are
branched. Every integral along  $C_{u_s}$ is
reduced to the integral along the $C_{v_s}$ contour by the
$z'\rightarrow g_s(z')$ transformation. As the result, in the
integrand, either $[y(z')-y(g_s(z'))(s_sz'+d_s)^2]$ or
$[\xi(z')-\xi(g_s(z'))(c_sz'+d_s)]$ appears. We replace every
above value by its value given by eqs.  (\ref{main1}) and
(\ref{main2}).  Eqs. (\ref{main1})-(\ref{main2}) are used also
to calculate both $[\xi(z')-\xi^{(s)}(z')]$ and
$[y(z')-y^{(s)}(z')]$ in every integral along the $\tilde C_s$
cut. The desirable equations turn out to be
\begin{eqnarray}
\xi(z)=
\sum_{s=1}^p\int\limits_{\tilde C_s}
G_{gh}^{(f)}(z,z')\left[[1-(-1)^{2l_{1s}}]
\varepsilon_s(z')
+\frac{\hat\epsilon_s(f,l_{1s})}{\sqrt{f'(z')}}
-\eta_s^{(ba)}(z')\right]\frac{dz'}{2\pi i}
\nonumber\\
+\sum_{s=p+1}^n\int\limits_{\tilde C_s}G_{gh}^{(f)}(z,z')
\left[[1-(-1)^{2l_{1s}}]\left(
\varepsilon_s(z')-
\frac{\hat\varepsilon_s(f)}{\sqrt{f'(z')}}\right)
-\eta_s^{(aa)}(z')\right]\frac{dz'}{2\pi i}
\nonumber \\
+\sum_{s=1}^p\int\limits_{C_{v_s}}
G_{gh}^{(f)}(z,z')\left[\epsilon_s(z',l_{2s})+
\frac{[1-(-1)^{2l_{2s}}]\hat\varepsilon_s(f)}{\sqrt{f'(z')}}
+\eta_s^{(ab)}(z')\right]\frac{dz'}{2\pi i}
\nonumber\\
+\sum_{s=p+1}^n\int
\limits_{C_{v_s}} G_{gh}^{(f)}(z,z')\left[\epsilon_s(z',l_{2s})
-\frac{\hat\epsilon_s(f,l_{2s})}{\sqrt{f'(z')}}
+\eta_s^{(bb)}(z')\right]
\frac{dz'}{2\pi i}
\label{eqxi}
\end{eqnarray}
together with the following ones
\begin{eqnarray}
y(z)=
\sum_{s=1}^p\int\limits_{\tilde C_s}G_{gh}^{(b)}(z,z')
\left[\rho_s^{(ba)}(z',l_{1s})-\frac{h_s(f)}{{f_0^{(s)}}'(z')}
\right]\frac{dz'}{2\pi i}+
\sum_{s=p+1}^n\int\limits_{\tilde C_s}G_{gh}^{(b)}(z,z')
\rho_s^{(aa)}(z')\frac{dz'}{2\pi i}
\nonumber \\
+\sum_{s=p+1}^n\int \limits_{C_{v_s}}
G_{gh}^{(b)}(z,z')\left[\frac{h_s(f)}{f_0'(g_s)g_s'(z')}-
\rho_s^{(bb)}(z)\right]\frac{dz'}{2\pi i}-
\sum_{s=1}^p\int\limits_{C_{v_s}}
G_{gh}^{(b)}(z,z')\rho_s^{(ab)}(z')\frac{dz'}{2\pi i}.
\label{eqy}
\end{eqnarray}
In (\ref{eqxi}) and (\ref{eqy}) the definition are the same as
in (\ref{main1}) and (\ref{main2}). Both $\eta^{(qr)}(z')$ and
$\rho^{(qr)}(z')$ are defined in Appendix A. Every $\tilde
C_s$ path in (\ref{eqxi}) and (\ref{eqy}) goes along the upper
edge of the cut from the $z_s^{(-)}$ point to the $z_s^{(+)}$
point where $z_s^{(+)}=g_s(z_s^{(-)})$ and $z_s^{(-)}$ is an
arbitrary point on the $C_{v_s}$ circle. Every $C_{v_s}$ circle
(\ref{circ}) in (\ref{eqxi}) and (\ref{eqy}) is rounded in the
positive direction starting from the $z_s^{(-)}$ point above.

Generally, the right
sides of (\ref{eqxi})-(\ref{eqy}) has a singularity at
$z=z_s^{(-)}$ and at $z=z_s^{(+)}$, as well. At the same time,
both $y(z)$ and $\xi(z)$ are assumed to have singularity
neither at the boundary of the fundamental domain $\Omega$ nor
inside $\Omega$. So, generally, the left sides of
(\ref{main1})-(\ref{main2}) being calculated from
(\ref{eqxi})-(\ref{eqy}), differ
from those given on the right side of above eqs.
(\ref{main1})-(\ref{main2}).  The additional contributions to
the left sides of eqs.(\ref{main1})-(\ref{main2}) are caused
by the proportional to zero mode terms on the right side of
(\ref{chgr}). Therefore, eqs.  (\ref{eqxi})-(\ref{eqy}) are
equivalent to (\ref{main1})-(\ref{main2}) only if the discussed
contributions are equal to zero. Hence eqs.
(\ref{eqxi}) and (\ref{eqy}) must be  added by the
following relations
\begin{eqnarray}
0=
\sum_{s=1}^p\int\limits_{\tilde C_s}
\tilde\chi_{F_r}(z')\left[[1-(-1)^{2l_{1s}}]
\varepsilon_s(z')
+\frac{\hat\epsilon_s(f,l_{1s})}{\sqrt{f'(z')}}
-\eta_s^{(ba)}(z')\right]\frac{dz'}{2\pi i}
\nonumber\\
+\sum_{s=p+1}^n\int\limits_{\tilde C_s}\tilde\chi_{F_r}(z')
\left[[1-(-1)^{2l_{1s}}]\left(
\varepsilon_s(z')-
\frac{\hat\varepsilon_s(f)}{\sqrt{f'(z')}}\right)
-\eta_s^{(aa)}(z')\right]\frac{dz'}{2\pi i}
\nonumber \\
+\sum_{s=1}^p\int\limits_{C_{v_s}}
\tilde\chi_{F_r}(z')\left[\epsilon_s(z',l_{2s})+
\frac{[1-(-1)^{2l_{2s}}]\hat\varepsilon_s(f)}{\sqrt{f'(z')}}
+\eta_s^{(ab)}(z')\right]\frac{dz'}{2\pi i}
\nonumber\\
+\sum_{s=p+1}^n\int
\limits_{C_{v_s}} \tilde\chi_{F_r}(z')\left[\epsilon_s(z',l_{2s})
-\frac{\hat\epsilon_s(f,l_{2s})}{\sqrt{f'(z')}}
+\eta_s^{(bb)}(z')\right]
\frac{dz'}{2\pi i}
\label{eqxich}
\end{eqnarray}
where $r=1,...n$, and to the following ones
\begin{eqnarray}
0=
\sum_{s=1}^p\int\limits_{\tilde C_s}\tilde\chi_{R_r}(z')
\left[\frac{h_s(f)}{{f_0^{(s)}}'(z')}-
\rho_s^{(ba)}(z',l_{1s})\right]\frac{dz'}{2\pi i}+
\sum_{s=p+1}^n\int\limits_{\tilde C_s}\tilde\chi_{R_r}(z')
\rho_s^{(aa)}(z')\frac{dz'}{2\pi i}
\nonumber \\
-\sum_{s=p+1}^n\int \limits_{C_{v_s}}
\tilde\chi_{R_r}(z')\left[\frac{h_s(f)}{f_0'(g_s)g_s'(z')}-
\rho_s^{(bb)}(z)\right]\frac{dz'}{2\pi i}+
\sum_{s=1}^p\int\limits_{C_{v_s}}
\tilde\chi_{R_r}(z')\rho_s^{(ab)}(z')\frac{dz'}{2\pi i}.
\label{eqych}
\end{eqnarray}
Eqs. (\ref{eqxi})-(\ref{eqych}) determine both $\xi(z)$ and
$y(z)$, as well as both  $(\hat\mu_s,\hat\nu_s)$ and
the $(\delta k_s,\delta u_s,\delta v_s)$
differences defined to be
\begin{equation}
\delta k_s=\hat
k_s-k_s^{(0)},\quad \delta u_s=\hat u_s-u_s^{(0)} \quad {\rm
and}\quad \delta v_s=\hat v_s-v_s^{(0)}.
\label{difsch}
\end{equation}
The $(k_s^{(0)},u_s^{(0)},v_s^{(0)})$ Schottky parameters in
(\ref{difsch}) are associated with the $g_s^{(0)}(f)$. A
various choice of the shape of the $\tilde C_s$ lines is
associated with various supermodular transformations.  Solving
(\ref{eqxi})-(\ref{eqych}), it is useful to use the following
relations \cite{dan5}
\begin{eqnarray}
\int\limits_{C_{v_s}} G_{gh}^{(f)}(z,z')P_{F_s}(z')
\frac{dz'}{2\pi i}-[1-(-1)^{2l_{1s}}]\int\limits_{\tilde C_s}
G_{gh}^{(f)}(z,z')P_{F_s}^{(a)}(z') \frac{dz'}{4\pi i}=0,
\nonumber\\
\int\limits_{C_{v_s}} G_{gh}^{(b)}(z,z')P_{R_s}(z')
\frac{dz'}{2\pi i}=0
\label{oo}
\end{eqnarray}
where $P_{R_s}(z')$, $P_{F_s}(z')$  and $P_{F_s}^{(a)}(z')$ are
defined by (\ref{poly}) and (\ref{pola}). To prove (\ref{oo}) one
can start with integrating a suitable Green function product
taken along the infinitesimal contour around $z$-point. Deforming
this contour to that, which  surrounds both $C_{v_s}$
and $C_{u_s}$ circles (\ref{circ}) together with the $\tilde
C_s$ cut ( if it exists), one uses (\ref{chgr}) and
(\ref{rgr}).  In more details the proof of the relations similar
to (\ref{oo}) is discussed in \cite{dan5}. Particular, owing to
(\ref{oo}), the sum of the integrals of $\varepsilon_s$ and
$\epsilon_s$ disappears in eq.(\ref{eqxi}). Furthermore, using
(\ref{norm}), (\ref{chgr}) and (\ref{oo}),  one
derives the following relations \cite{dan5}
\begin{eqnarray}
\int\limits_{C_{v_s}} \tilde\chi_{F_r'}(z')P_{F_s}(z')
\frac{dz'}{2\pi i}-[1-(-1)^{2l_{1s}}]\int\limits_{\tilde C_s}
\tilde\chi_{F_r'}(z')P_{F_s}^{(a)}(z') \frac{dz'}{4\pi
i}=-\delta_{F_r'F_s},
\nonumber\\
\int\limits_{C_{v_s}}
\tilde\chi_{R_r'}(z')P_{R_s}(z') \frac{dz'}{2\pi i}=
\delta_{R_r'R_s}.
\label{oochi}
\end{eqnarray}
where $\tilde\chi_{R_r}(z')$ and $\tilde\chi_{F_r}(z')$ are the
same as in (\ref{eqxich}) and in (\ref{eqych}), $\delta_{NN'}$
being the Kronecker symbol. Particular, using (\ref{oochi}),
one can perform explicitly the integrals of $\varepsilon_s$
and $\epsilon_s$ in (\ref{eqxich}). In this case (\ref{eqxich})
are written down as
\begin{eqnarray}
\hat\mu_r=\sum_{s=1}^n\left[\hat X_{\mu_r\mu_s}^{-1}(L)\mu_s+
\hat X_{\mu_r\nu_s}^{-1}(L)\nu_s\right]+\sum_{s=1}^n
\hat X_{\mu_rF_s}^{-1}(L)\eta_{F_s}\,,
\nonumber\\
\hat\nu_r=\sum_{s=1}^n\left[\hat X_{\nu_r\mu_s}^{-1}(L)\mu_s+
\hat X_{\nu_r\nu_s}^{-1}(L)\nu_s\right]+\sum_{s=1}^n
\hat X_{\nu_rF_s}^{-1}(L)\eta_{F_s}
\label{eqodd}
\end{eqnarray}
where $\eta_{F_r}$ is defined as
\begin{eqnarray}
\eta_{F_r}=2\sum_{s=1}^p\int\limits_{C_{v_s}}
\tilde\chi_{F_r}(z)\eta_s^{(ab)}(z)\frac{dz}{2\pi i}
+2\sum_{s=p+1}^n\int\limits_{C_{v_s}} \tilde\chi_{F_r}(z)
\eta_s^{(bb)}(z)\frac{dz}{2\pi i}
\nonumber\\
-2\sum_{s=1}^p\int\limits_{\tilde C_s}
\tilde\chi_{F_r}(z)\eta_s^{(ba)}(z)\frac{dz}{2\pi i}
-2\sum_{s=p+1}^n\int\limits_{\tilde C_s}\tilde\chi_{F_r}(z)
\eta_s^{(aa)}(z)\frac{dz}{2\pi i}\,.
\label{etafr}
\end{eqnarray}
The $\hat X_{F_rF_s}(L)$  matrix in (\ref{eqodd}) is given by
\begin{eqnarray}
\hat X_{F_r\mu_s}(L)=2\frac{1-(-1)^{2l_{2s}}}{\hat u_s-\hat v_s}
\int\limits_{C_{v_s}}dz\frac{\tilde\chi_{F_r}(z)
[f(z)-\hat v_s]}{2\pi i\sqrt{f'(z)}}\nonumber \\
-2\frac{(-1)^{2l_{1s}}+
\sqrt{\hat k_s}}{\hat u_s-\hat v_s}\int\limits_{\tilde C_s}dz
\frac{\tilde\chi_{F_r}(z)[f(z)-\hat v_s]}{2\pi i\sqrt{\hat
k_sf'(z)}} \quad{\rm for}\quad s\leq p;\nonumber \\
\hat X_{F_r\mu_s}(L)=2\frac{(-1)^{2l_{2s}}+\sqrt{\hat
k_s}}{\hat u_s-\hat v_s}
\int\limits_{C_{v_s}}dz\frac{\tilde\chi_{F_r}(z)
(f(z)-\hat v_s)}{2\pi i\sqrt{\hat k_sf'(z)}}\nonumber \\
-2\frac{1-(-1)^{2l_{1s}}}{\hat u_s-\hat v_s}
\int\limits_{\tilde
C_s}dz\frac{\tilde\chi_{F_r}(z) [f(z)-\hat v_s]}{2\pi
i\sqrt{f'(z)}}
\quad{\rm for}\quad s\geq p; \nonumber \\
\hat X_{F_r\nu_s}(L)=2\frac{(-1)^{2l_{2s}}-1}{\hat u_s-\hat v_s}
\int\limits_{C_{v_s}}dz\frac{\tilde\chi_{F_r}(z)
[f(z)-\hat v_s]}{2\pi i\sqrt{f'(z)}}\nonumber \\
+2\frac{(-1)^{2l_{1s}}\sqrt{\hat k_s}+
1}{\hat u_s-\hat v_s}\int\limits_{\tilde C_s}dz
\frac{\tilde\chi_{F_r}(z)[f(z)-\hat v_s]}{2\pi i\sqrt{
f'(z)}} \quad{\rm for}\quad s\leq p;\nonumber \\
\hat X_{F_r\nu_s}(L)=-2\frac{(-1)^{2l_{2s}}\sqrt{\hat
k_s}+1}{\hat u_s-\hat v_s}
\int\limits_{C_{v_s}}dz\frac{\tilde\chi_{F_r}(z)
(f(z)-\hat v_s)}{2\pi i\sqrt{f'(z)}}\nonumber \\
+2\frac{1-(-1)^{2l_{1s}}}{\hat u_s-\hat v_s}
\int\limits_{\tilde
C_s}dz\frac{\tilde\chi_{F_r}(z) [f(z)-\hat v_s]}{2\pi
i\sqrt{f'(z)}}
\quad{\rm for}\quad s\geq p;
\label{hatX}
\end{eqnarray}
Since the terms on the right sides of (\ref{eqxi}), (\ref{eqy}),
(\ref{eqych}) and (\ref{eqodd}) are proportional to Grassmann
parameters, the above equations can be solved by an iteration
procedure. In this case the solution of the considered equations
is obtained to be series in $\{\mu_s,\nu_s\}$.
The first step of the discussed procedure is to replace $f(z')$
in (\ref{eqxi}) and in (\ref{eqodd}) by $f_0(z')$ (assumed to be
known), the terms with $\eta^{(qr)}$ ( $q=a,b$ and $r=a,b$ )
being neglected. In this case (\ref{eqxi}) determines the linear
in $\{\mu_s,\nu_s\}$ terms in $\xi(z)$. In the same approximation
eqs. (\ref{eqodd}) determine the action of the supermodular group
on the odd super-Schottky parameters. It is obvious that in both
$y(z)$ and the $(\delta k_s,\delta u_s,\delta v_s)$ differences,
the dependence on odd parameters begins with quadratic terms.
These terms are calculated by the substitution of the obtained
$\xi(z)$, $\hat\mu_s$ and $\hat\nu_s$ to (\ref{eqy}) and to
(\ref{eqych}), $f(z')$ being replaced by $f_0(z')$. One employs
also that in the considered approximation, $h_s(z)$ is given by
\begin{equation}
h_s(z)=P_{k_s}^{(0)}\delta k_s+P_{u_s}^{(0)}\delta u_s+
P_{v_s}^{(0)}\delta v_s
\label{hdel}
\end{equation}
where $P_{k_s}^{(0)}$, $P_{u_s}^{(0)}$ and $P_{v_s}^{(0)}$
are the $P_{k_s}$, $P_{u_s}$ and $P_{v_s}$ polynomials in
(\ref{poly}) associated with the $g_s^{(0)}(z)$
transformations.  The iteration procedure being continued, one
calculates $\xi(z)$, $y(z)$ and in $(\delta k_s,\delta
u_s,\delta v_s)$ differences to be series
over odd modular parameters. One can prove that both $\xi(z)$
and $y(z)$ have no singularities  inside the fundamental domain
$\Omega$ and at its boundary, as well ( the proof is omitted
here). Simultaneously, all the resulted values in question
depend on the $L$ superspin structure, if the
$\{k_s,u_s,v_s,\mu_s,\nu_s\}$ set is chosen to be the same for
all the superspin structures. The above $L$
dependence arises because both $G_{gh}^{(f)}(z,z')$ in
(\ref{eqxi}) and $\tilde\chi_F(z')$ in (\ref{eqxich}) depend on
$L$. One can see also that  all the even
superspin structures $S_{ev}$ without odd genus-1 structures can
be obtained by a suitable supermodular transformation of the
only superspin structure. In the next Section we consider the
supermodular transformations turning the above $S_{ev}$
structures to those containing an even number of the odd
genus-1 spinor structures.

\section{Transformation of two even genus-1 structures
to a pair of the odd genus-1 ones}

In this Section we consider the
supermodular transformations of $S_{ev}\rightarrow S_{2}$, which
turn a pair of the genus-1 structures to a pair of the odd
genus-1 ones, say
$l_{1s_1}=l_{1s_2}=1/2,l_{2s_1}=l_{2s_2}=0\rightarrow
l_{1s_1}=l_{1s_2}=l_{2s_1}=l_{2s_2}=1/2$. Without loss of
generality we assume that $s_1=1$ and $s_2=2$.  Under the
discussed transformation, the $\omega_{12}^{(r)}$ element of the
$\omega^{(r)}$ period matrix turns to $\omega_{12}^{(r)}\pm1$,
the other $\omega^{(r)}$ matrix elements being unchanged. It is
worth while to note that the expression \cite{martnp,fried,vec7}
of $\omega_{12}^{(r)}$ contains, among of other things, the
following term
\begin{equation}
\omega_{12}^{(r)}=\frac{1}{2\pi
i}\ln\frac{(u_1-u_2)(v_1-v_2)}{(u_1-v_2)(v_1-u_2)}+...
\label{om12}
\end{equation}
The desirable $\omega_{12}^{(r)}\rightarrow
\omega_{12}^{(r)}\pm1$ replacement is achieved by the addition
of $2\pi$ to the argument of one of the differences inside the
round brackets in (\ref{om12}). So the discussed transformation
of $\omega$ appears to be the result of suitable rounds of the
$(u_1,u_2,v_1,v_2)$ fixed points over each other.  As example,
we consider the clock-wise going of the $u_2$ point about the
$u_1$ one. On $z$-plane the discussed round corresponds to the
going of $C_{u_2}$ circle (\ref{circ}) about $C_{u_1}$
circle, as it is shown in Fig.1. In this case we start with the
$S_{ev}$ superspin structure with
$l_{11}=l_{12}=1/2,l_{21}=l_{22}=0$. The initial position of the
circles and the cuts is shown in Fig.1(a). The $(v_1)(u_1)$  cut
is situated between $v_1$ and $u_1$ fixed points
and the $(v_2)(u_2)$  cut is situated between $v_2$ and $u_2$.
After the discussed round to be performed, the circles are
returned again to the same position, but the cuts are deformed,
as it is shown in Fig.1(b).  A shape of the cuts is unessential
because, as it has been explained in Section 2, it can be
changed by a supermodular transformation. Hence we can close the
cuts together, as it is shown in Fig.1(c). In Fig.1(c) the
resulted cuts are represented by the bold lines. As it is
follows from Fig.1(b), the bold line about $C_{u_1}$ circle is
formed by the $(v_2)(u_2)$ cut sandwiched by $(v_2)(u_2)$
cut. The bold line about $C_{u_2}$ circle is
formed by the $(v_2)(u_2)$ cut. And the bold line drawing from
$C_{u_1}$ circle to $C_{u_2}$ one, is formed by both
$(v_1)(u_1)$ and $(v_2)(u_2)$ cuts  sandwiched  every other.
The cuts surrounding both the circles in
Fig.1(c) are removed by the re-definition of superholomorphic
functions at ( and inside ) the considered circles to be
the analytical continuation of the above superholomorphic
functions from the fundamental domain $\Omega$.  As we noted
already in Section 2, the $\Omega$ domain presents the exterior
of all the $C_{v_s}$ circles and of the $C_{u_s}$ ones. Under
the above re-definition, both the $(\Gamma_{b,s}(l_{2s}=0)$
transformations for both $s=1$ and $s=2$ are also changed to be
$\Gamma_{b,s}^{(ch)}$ as follows
\begin{equation}
\Gamma_{b,s}^{(ch)}=
\delta_{s1}\Gamma_{a,1}\Gamma_{a,2}\Gamma_{a,1}
\Gamma_{b,1}(l_2=0)+\delta_{s2}
\Gamma_{a,1}\Gamma_{b,2}(l_2=0)\quad{\rm for}\quad s=1,2
\label{tg}
\end{equation}
In eq.(\ref{tg}), the $\Gamma_{a,s}\equiv
\Gamma_{a,s}(l_{1s}=1/2)$ mappings are defined by
(\ref{gammaa}) and $\delta_{sr}$ is the Kronecker symbol. As it
follows from (\ref{schot}) and (\ref{around}), in the case of
the odd modular parameters to be zero ( $\mu_1=\nu_1=
\mu_2=\nu_2=0$ ), the transformation (\ref{tg}) just corresponds
to the transformation of two considered even genus-1 spinor
structures into the odd genus-1 spinor structures:
$l_{21}=l_{22}=0\rightarrow l_{21}=l_{22}=1/2$,
$l_{11}=l_{12}=1/2$ being unchanged. For arbitrary modular
parameters, every $T_p(t)$ superconformal $p$-tensor is changed
in going across the $(u_1)(u_2)$ line in Fig.1(c) by the
$\Gamma_{2121}$ transformation turned out to be
\begin{equation}
\Gamma_{2121}=\Gamma_{a,2}\Gamma_{a,1}\Gamma_{a,2}\Gamma_{a,1}
\label{gcut}
\end{equation}
where $\Gamma_{a,s}$ are the same as in (\ref{tg}).
Eq.(\ref{gcut}) follows directly from (\ref{stens})
and Fig.1(b). Hence in this case the cut arises to be
between $C_{u_1}$ and $C_{u_2}$.
Nevertheless, we show that this cut can be removed by a
suitable superholomorphic mapping of the $t=(z|\theta)$
supercoordinate.  As the result, we again obtain two odd genus-1
structures. We write down the desired
mapping $t\rightarrow\tilde t=(\tilde z|\tilde\theta)$ of the $t$
supercoordinate as  follows
\begin{equation}
z=\tilde f(\tilde z)+\tilde f'(\tilde
z)\tilde\theta\tilde\xi(\tilde z)\quad{\rm and}\quad
\theta=\sqrt{\tilde f'(\tilde z)}\left[\left(1+
\frac{1}{2}\tilde\xi(\tilde z)\tilde\xi'(\tilde z)\right)
\tilde\theta+\tilde\xi(\tilde z)\right]
\label{smtrodd}
\end{equation}
where both $\tilde f(\tilde z)$ and $\tilde\xi(\tilde z)$ are
proportional to the odd modular parameters. In this case
the $\{\tilde u_s,\tilde v_s,\tilde k_s\}$ resulted
Schottky parameters differ from the old Schottky ones only by
terms proportional to the odd modular parameters. Furthermore,
$A_s\rightarrow A_s$ and $B_s\rightarrow B_s$ under the above
mapping (\ref{smtrodd}). Like the previous Section, we calculate
that solution of (\ref{smtrodd}), which does not change the
$\{N_0\}$ set of $(3|2)$ Schottky parameters chosen to be no
moduli. Both $\tilde f(\tilde z)$ and $\tilde\xi(\tilde z)$ are
calculated from the condition that superconformal tensors  have
no discontinuity going across the $(\tilde u_1)(\tilde u_2)$
line on $\tilde z$-plane. In this case
\begin{eqnarray}
\tilde\xi^{(l)}(\tilde z)-\tilde\xi^{(r)}(\tilde z)=
4[\varepsilon_2(\tilde f)-\varepsilon_1(\tilde f)]-12
\varepsilon_1(\tilde f)\varepsilon_2(\tilde
f)\varepsilon_2'(\tilde f)+4\varepsilon_1(\tilde f)
\varepsilon_2(\tilde f)\varepsilon_1'(\tilde f)
\nonumber\\
{\rm and}\quad
\tilde f^{(l)}(\tilde z)-\tilde f^{(r)}(\tilde z)=
-8\varepsilon_1(\tilde f)\varepsilon_2(\tilde f)
\label{dis}
\end{eqnarray}
where $\tilde f\equiv \tilde f(\tilde z)$ and the $(l)$ symbol
( and, respectively, the $(r)$ symbol ) at
the right top shows that the value being marked by the above
symbol, is calculated at the left ( and, respectively, the right
) edge of the $(\tilde u_1)(\tilde u_2)$ cut. Eqs.(\ref{dis})
follow directly from (\ref{gcut}). Above eqs.(\ref{dis}) must be
complected by the relations
\begin{eqnarray}
\Gamma_{b,s}^{(ch)}(t)=
t\left(\tilde\Gamma_{b,s}(l_{2s})(\tilde t)\right)\quad {\rm
and}\quad \Gamma_{a,s}(l_{1s})(\hat t)= t^{(s)}
(\left(\tilde\Gamma_{a,s}(l_{1s})(\tilde t)\right)) \quad {\rm
for}\quad s=1,2;
\nonumber\\
\Gamma_{b,s}(l_{2s})(t)=
t\left(\tilde\Gamma_{b,s}(l_{2s})(\tilde t)\right)\quad {\rm
and}\quad \Gamma_{a,s}(l_{1s})(t)=
t^{(s)}\left(\tilde\Gamma_{a,s}(l_{1s})(\tilde t)\right)\quad
{\rm for}\quad s>2.
\label{mn1}
\end{eqnarray}
where $(\tilde\Gamma_{a,s}(l_{1s}),\tilde\Gamma_{b,s}(l_{2s}))$
transformations are associated with $2\pi$-twists about
$(A_s,B_s)$-cycles on $\tilde t$-supermanifold. Furthermore,
$l_{1s}=l_{2s}=1/2$ for $s=1,2$ and $\Gamma_{b,s}^{(ch)}$ are
given by (\ref{tg}). Eqs.(\ref{mn1}) determine the
discontinuities of both $\tilde f(\tilde z)$ and
$\tilde\xi(\tilde z)$ under $2\pi$-twists about
$(A_s,B_s)$-cycles. Eqs.(\ref{dis})-(\ref{mn1}) can be solved
by the same method, which was developed in the previous Section
for the solution of eqs.(\ref{main}). In this case both  $\tilde
f(\tilde z)$ and $\tilde\xi(\tilde z)$ are found to be quite
similar to $f(z)$ and $\xi(z)$ in Section 3 except only the
additional term due to eqs.(\ref{dis}). To avoid unwieldy
expressions we give in the  explicit form only the terms linear
in odd modular parameters. In this approximation, $\tilde
f(\tilde z)=\tilde z$.  In addition, there is no difference
between $\tilde\xi(\tilde z)$ and $\tilde\xi(z)$. In the
considered approximation, the desired $\tilde\xi(z)$ turns out
to be
\begin{equation}
\tilde\xi(z)=-\int\limits_{\tilde
C_{12}}G_{gh}^{(f)}(z,z')\gamma(z')\frac{dz'}{2\pi
i}+\int\limits_{C_{u_1}}G_{gh}^{(f)}(z,z')\gamma(z')
\frac{dz'}{4\pi i}-
\int\limits_{C_{u_2}}G_{gh}^{(f)}(z,z')\gamma(z')
\frac{dz'}{4\pi i}
\label{hatxi}
\end{equation}
In (\ref{hatxi}) the $\tilde C_{12}$ path goes along the $u_1u_2$
cut in Fig.1(c) from the $z_1^{(+)}$ point to the $z_2^{(+)}$
one. Every $z_s^{(+)}$ point  ( $s=1,2$ ) is the
intersect of the $\tilde C_{12}$ path with the
$C_{u_s}$ circle. Every $C_{u_s}$-circle ( $s=1,2$ ) in
(\ref{hatxi}) is rounded in the positive direction starting from
the $z_s^{(-)}$ point defined above. The $\gamma(z)$
local parameter in (\ref{hatxi}) is given by
\begin{equation}
\gamma\equiv\gamma(z)=
4\varepsilon_2(z)-4\varepsilon_1(z)
\label{ga}
\end{equation}
where $\varepsilon_s(z)$ are defined by (\ref{eps}) for
$s=1,2$. In deriving (\ref{hatxi}) we transform the integrals
along the $C_{v_s}$ circles to the integrals along the $C_{u_s}$
circles by the $z\rightarrow (d_sz-b_s)(-c_sz+a_s)^{-1}$
replacements.  In addition, (\ref{hatxi}), eqs.(\ref{oo}) have
been taken into account. Using (\ref{eps}) and (\ref{pola}) one
can express $\gamma(z)$ in terms of $P_{F_s}^{(a)}$ polynomials
as follows
\begin{equation}
\gamma(z)=\mu_2 P_{\mu_2}^{(a)}(z)+ \nu_2 P_{\nu_2}^{(a)}(z)
-\mu_1 P_{\mu_1}^{(a)}(z)-\nu_1 P_{\nu_1}^{(a)}(z)
\label{gap}
\end{equation}
One can verify that $\tilde\xi(z)$ being
given by (\ref{hatxi}), satisfies eq.(\ref{dis}) taken in the
considered linear approximation in odd modular parameters. In
addition, (\ref{hatxi}) has no singularities at $z=z_1^{(+)}$ and
at $z=z_2^{(+)}$. Since the dependence on Grassmann
parameters of the even moduli begins with quadratic terms, only
the $(\mu_s,\nu_s)$ Schottky parameters are changed in the
considered linear approximation in odd modular parameters, the
transformed parameters being $(\tilde\mu_s,\tilde\nu_s)$. The
$(\tilde\mu_s,\tilde\nu_s)$ in question are calculated from the
relations quite similar to (\ref{eqxich}). Eqs.(\ref{oochi})
being taken into account, the desired
$(\tilde\mu_s,\tilde\nu_s)$ are found to be
\begin{equation}
\tilde\mu_s=\mu_s+
\sum_{j=1}^2X_{\mu_s\mu_j}\mu_j+ X_{\mu_s\nu_j}\nu_j
\quad{\rm and}\quad
\tilde\nu_s=\nu_s+\sum_{j=1}^2X_{\nu_s\mu_j}\mu_j+
X_{\nu_s\nu_j}\nu_j
\label{munu}
\end{equation}
where $X_{\mu_s\mu_j}$, $X_{\mu_s\nu_j}$, $X_{\nu_s\mu_j}$
and $X_{\nu_s\nu_j}$ are the non-zero matrix elements
$X_{F_sF_j'}$ of the $X$ matrix, which is defined by
\begin{eqnarray}
(-1)^jX_{F_sF_j'}=\int\limits_{C_{u_1}}\tilde\chi_{F_s}(z')
P_{F_j'}^{(a)}(z')\frac{dz'}{2\pi i}
-2\int\limits_{\tilde
C_{12}}\tilde\chi_{F_s}(z')P_{F_j'}^{(a)}(z')\frac{dz'}{2\pi i}
-\int\limits_{C_{u_2}}\tilde\chi_{F_s}(z')P_{F_j'}^{(a)}(z')
\frac{dz'}{2\pi i}
\nonumber \\
{\rm  for}\quad F_j'=\mu_j,\nu_j \quad{\rm with}\quad j=1,2;
\nonumber \\
X_{F_sF_j'}=0\quad
{\rm  for}\quad F_j'=\mu_j,\nu_j \quad{\rm with}\quad j>2.
\label{res}
\end{eqnarray}
where $X_{F_sF_j'}$ matrix elements are
labeled by $F_s=(\mu_s,\nu_s)$ and $F_j'=(\mu_j,\nu_j)$ where
$1\leq s\leq n$, $1\leq j\leq n$, $n$ being the genus. In
deriving (\ref{res}) eq.(\ref{gap}) is used. It is worth-while
to note that $X_{F_sF_j'}=0$, if $F_s\in\{N_0\}$. In the next
section we employ eqs.(\ref{res}) to discuss the supermodular
covariance of the multi-loop partition functions.

\section{Supermodular covariance of the superstring partition
functions in the particular case}

It is commonly to write n-loop superstring amplitudes $A_n$ as
follows
\begin{equation}
A_n=\int\prod_N dq_Nd\overline q_N\prod_r dt^{(r)}d\overline
t^{(r)} \sum_{L,L'}\hat Z_{L,L'}(\{q_N,\overline q_N\})
<\prod_rV(t^{(r)},\overline t^{(r)})>_{L,L'}
\label{ampl}
\end{equation}
where $\hat Z_{L,L'}$ are the measures ( partition functions )
and the $<...>_{L,L'}$ symbol denotes the vacuum expectations
calculated for the $(L,L')$ superspin structure. The index $L$
($L'$) labels superspin structures of right (left) fields. The
integration in (\ref{ampl}) is performed  over both
$(3n-3|2n-2)$ complex moduli $q_N$ and over their complex
conjugated $\overline q_N$ and, in addition, over the
$(z^{(r)},\overline z^{(r)})$ vertex local coordinates and over
their odd partners $(\theta^{(r)},\overline\theta^{(r)})$, as
well. As it was already noted in Section 1, in fact
eq.(\ref{ampl}) needs the regularization. In this Section we
employ eq.(\ref{ampl}) only to clean the definition of the
$\hat Z_{L,L'}$ partition functions.  The holomorphic structure
\cite{pst,cqg} of the above partition functions is determined by
the following equation
\begin{equation}
\hat Z_{L,L'}(\{q_N,\overline q_N\})=\left[\det2\pi
i[\overline{\omega(\{q_{N_s}\},L')}-
\omega(\{q_{N_s}\},L)]\right]^{-5}
Z_L(\{q_{N_s}\})
\overline {Z_{L'}(\{q_{N_s}\})}
\label{hol}
\end{equation}
where $Z_L(\{q_{N_s}\})$ is a holomorphic function of the
$q_{N_s}$ moduli, and $\omega(\{q_{N_s}\},L)$ is the period
matrix associated with the supermanifold under consideration.
In terms of super-Schottky group parameters
$\{k_s,u_s,v_s,\mu_s,\nu_s\}$ both $Z_L(\{q_{N_s}\})$ and
$\omega(\{q_{N_s}\},L)$ for the Neveu-Schwarz sector have been
obtained in \cite{vec8,pst,dan1,dan0}. For all the even superspin
structures in the Ramond sector they have been
calculated in \cite{dan5}. In this case the result is given in
the form of series over Grassmann modular parameters.  For the
sake of completeness we present these results in Appendix B.

It is necessary for the considered theory to be self-consistent,
that the above $\hat Z_{L,L'}$ partition functions do be
covariant under $q_N\rightarrow\hat q_N(\{q_N\})$ supermodular
group transformations of the $q_N$ modular parameters as follows
\begin{equation}
\hat Z_{\hat L,\hat L'}(\{\hat q_N,\overline{\hat q_N}\})=
\hat Z_{L,L'}(\{q_N,\overline q_N\})
|Jac(\partial q_N/\partial\hat q_{N'})|^2
\label{sdet}
\end{equation}
where $Jac(\partial q_N/\partial \hat q_{N'})$ is the Jacobian
of the considered supermodular transformation and $\hat L$
$(\hat L')$ is the resulted superspin structure of right (left)
fields.  We give a direct evidence that, for zero odd moduli,
the partition function calculated in \cite{dan5} satisfy
eq.(\ref{sdet}) under the supermodular transformations
turning a pair of the genus-1 structures to a pair of the odd
genus-1 ones. In the next Section we argue that the considered
partition functions are covariant under the whole supermodular
group.

At zero odd moduli the $\omega(\{q_{N_s}\},L)$ period matrix in
(\ref{hol}) is reduced to that associated with the Riemann
surface, and, therefore, it is independent of the $L$ spin
structure. In this case, as it follows from (\ref{sdet}) and
from (\ref{hol}), the $Z_{S_{ev}}(\{q_{N_s}\})$ holomorphic
partition function assigned to the $S_{ev}$ spin structure is
changed under the $S_{ev}\rightarrow S_2$ supermodular
transformation discussed in Section 4, as follows
\begin{equation}
Z_{S_{ev}}(\{q_N\})=\frac{ Z_{S_2}(\{\tilde q_N\})}{\det(I+X)}
\label{zerap}
\end{equation}
where $Z_{S_2}$ is the partition function associated with the
$S_2$ spin structure. The above ($S_{ev},S_2$) spin structures
were defined in the end of Section 3 and in the beginning of the
previous Section. The $X$ matrix in (\ref{zerap}) is defined
by (\ref{res}). We show that the partition functions obtained in
\cite{dan5} satisfy the conditions (\ref{zerap}). One can see
from Appendix B that discussed $Z_L$ partition functions
at zero odd modular parameters can be written down
as follows
\begin{equation}
Z_L(\{k_s,u_s,v_s\})=
\frac{\tilde Z_L(\{k_s,u_s,v_s\},\{\sigma_p\})}
{\sqrt{\det \tilde M(\{\sigma_p\})
\det \tilde M(\{-\sigma_p\})}}
\label{zgh}
\end{equation}
where $\tilde Z_L(\{k_s,u_s,v_s\})$ is invariant under the
discussed $S_{ev}\rightarrow S_2$ supermodular transformations
and $\tilde M(\{\sigma_p\})$ is the matrix defined below. Both
$\tilde M(\{\sigma_p\})$  and $\tilde
Z_L(\{k_s,u_s,v_s\},\{\sigma_p\})$  depend on the choice of the
$\{\sigma_p\}$ set where $\sigma_p=\pm1$, and $p$ labels those
genus-1 spin structures, which are associated with $l_{1p}=1/2$.
Nevertheless, the right side of (\ref{zgh}) turns out to be
independent of the choice of the above $\{\sigma_p\}$ set
\cite{dan5}. To define the $\tilde M(\{\sigma_p\})$ matrix in
(\ref{zgh}) we consider \cite{dan5} for every spin structure the
Green functions $G_{(\sigma)}(z,z')$ and the Green function
$G_f(z,z')$ as it follows just below.

The $G_{(\sigma)}(z,z')$  functions are defined \cite{dan5} by
\begin{equation}
G_{(\sigma)}(z,z')= \sum_\Gamma\frac{\exp\pi i
[\Omega_\Gamma(\{\sigma_s\})+\sum_s2l_{1s}
\sigma_s(J_{(o)s}(z)-J_{(o)s}(z'))]}
{[z-g_\Gamma(z')][c_\Gamma z'+d_\Gamma]^3}
\label{grsig}
\end{equation}
where $J_{(o)s}$ are the functions having the periods to be $2\pi
i\omega_{sr}^{(r)}$, and $\omega_{sr}^{(r)}$ is the period
matrix at zero odd moduli.  The summation in (\ref{grsig}) is
performed over all the group products $\Gamma=\{z\rightarrow
g_\Gamma(z)\}$ of the basic group elements
$\Gamma_s=\{z\rightarrow g_s(z)\}$ including $\Gamma=I$.
Furthermore, $\Omega_\Gamma(\{\sigma_s\})$ in (\ref{grsig}) is
defined as
\begin{equation}
\Omega_\Gamma(\{\sigma_s\})=-\sum_{s,r}2l_{1s}
\sigma_s\omega_{sr}^{(r)}n_r(\Gamma)+
\sum_r(2l_{2r}-1)n_r(\Gamma)
\label{omgm}
\end{equation}
where $n_r(\Gamma)$ is the
number of times that the $\Gamma_r$ generators  are present
in $\Gamma$ (for its inverse $n_r(\Gamma)$ is defined to be
negative ) and $\sigma_s=\pm1$. So, $G_{(\sigma)}$ depends on a
choice of the $\{\sigma_s\}$ set.
It is follows from (\ref{grsig}) that the changes of
$G_{(\sigma)}$ under the $z\rightarrow g_s(z)$ mappings are
as follows
\begin{equation}
G_{(\sigma)} (g_s(z),z')=(-1)^{2l_2-1}(c_rz+d_r)^{-1}
\left(G_{(\sigma)} (z,z')
+\sum_{F_r=\mu_r,\nu_r}\tilde Y_{\sigma,F_r}(z)
\Phi_{\sigma,{F_r}}^{(0)}(z')\right)
\label{chgsig}
\end{equation}
where  $\Phi_{\sigma,{F_r}}^{(0)}(z')$ are 3/2-tensors,
$F_r=(\mu_r,\nu_r)$ and
$\tilde Y_{\sigma,F_r}(z)$  is given by
\begin{equation}
\tilde
Y_{\sigma,F_r}(z)=\exp[\pi
i\sum\nolimits_s2l_{1s}\sigma_sJ_{(o)s}(z)] P_{F_r}(z)
\label{ysig}
\end{equation}
where $P_{F_r}(z)$ is given by (\ref{poly}). In addition, we
define the $G_f(z,z')$ Green function, which is changed under
$2\pi$-twists about $A_s$-cycles and about $B_s$-ones as follows
\begin{eqnarray}
G_f(g_s(z),z')=(-1)^{2l_{2s}-1}(c_sz+d_s)^{-1}\left(G_f(z,z')+
\sum_{F_s=\mu_s,\nu_s}P_{F_s}(z)\chi_{F_s}(z')\right)
\nonumber\\
G_f(z,z')=-\frac{[1-(-1)^{2l_{1_s}}]}{2}
\left(G_f^{(s)}(z,z')+
\sum_{F_s=\mu_s,\nu_s} P_{F_s}^{(a)}(z)\chi_{F_s}(z')\right)
\label{trzgr}
\end{eqnarray}
where, unlike (\ref{chgr}) and (\ref{rgr}), the summation is
performed over all
$F_s=(\mu_s,\nu_s)$ including the $\{N_0\}$ set, too.
FUrthermore,  $G_f^{(s)}(z,z')$ in (\ref{trzgr}) is the
$G_f$ Green function $2\pi$-twisted about $A_s$-cycle.  In
addition, $P_{F_s}(z)$ and $P_{F_s}^{(a)}(z)$ in (\ref{trzgr})
are degree-1 polynomials defined by (\ref{poly}) and
(\ref{pola}), $\chi_{F_s}(z')$ being conformal
3/2-tensors.\footnote{ In terms of the
$\chi_{N_s}(t')$ superconformal 3/2-tensors defined in
\cite{dan5}, every above $\chi_{F_s}(z')$ is equal to
$\chi_{F_s}(t')$ taken at zero odd modular parameters.} Above
$\chi_{F_s}(z')$ have no singularities in the fundamental domain
on $z'$-plane, except only at $z'\rightarrow\infty$. It is
worth-while to note that the $G_{gh}^{(f)}(z,z')$ Green
function discussed in the previous Sections can be expressed in
terms of the $G_f(z,z')$ function as
\begin{equation}
G_{gh}^{(f)}(z,z')=G_f(z,z')-\sum_{F\in\{N_0\}}
\tilde P_F(z)\chi_F(z')
\label{ghgr}
\end{equation}
where $\tilde P_F(z)$ are degree-1 polynomials in $z$. The $F$
indices in (\ref{ghgr}) are associated with those odd Schottky
parameters, which chosen to be the same for all genus-$n$
supermanifolds.  The $\tilde P_F(z)$ polynomials in (\ref{ghgr})
are determined from the condition for $G_{gh}^{(f)}(z,z')$ to
decrease at least as $(z')^{-3}$ when $z'\rightarrow\infty$. One
can prove \cite{dan5} that, simultaneously, this condition
provides relations (\ref{chgr}). It
can be also proved \cite{dan5} that the $\tilde P_F(z)$
polynomials in (\ref{ghgr}) are independent of the spin
structure.

To define the desired $\tilde M(\{\sigma_p\})$ matrix in
(\ref{zgh}) it is worth-while to note that the above $G_f(z,z')$
functions can be expressed in terms $G_{(\sigma)} (z,z')$ as
\begin{eqnarray}
G_f(z,z')=
G_{(\sigma)} (z,z')-\sum_{s=1}^n\sum_{F_s=\mu_s,\nu_s}
\int\limits_{C_{v_s}}
G_{(\sigma)} (z,z'')\frac{ dz''} {2\pi i}
P_{F_s}(z'')\chi_{F_s}(z') \nonumber \\
+\sum_{s=1}^n\sum_{F_s=\mu_s,\nu_s}
\frac{[1-(-1)^{2l_{1_s}}]}{2}\int\limits_{\tilde C_s}
G_{(\sigma)} (z,z'')\frac{ dz''} {2\pi i}
P_{F_s}^{(a)}(z'')
\chi_{F_s}(z')
\label{gfgsig}
\end{eqnarray}
where $\chi_{F_s}(z')$ are
3/2-tensors defined by (\ref{trzgr}). Both the
$C_{v_s}$ contours and the $\tilde C_s$ ones are the
same as in (\ref{eqxi}) and (\ref{eqy}). Furthermore,
the $P_{F_s}(z'')$ polynomials are defined by (\ref{poly})
and the $P_{F_s}^{(a)}(z'')$ ones are given by (\ref{pola}). To
derive (\ref{gfgsig}) we represent $G_f(z,z')$ to be the
integral over $z''$ performed along the infinitesimal contour
around $z'$, the integrand being $G_{(\sigma)} (z,z'')$.
Running this contour away and using (\ref{chgr}) and
(\ref{rgr}), we obtain (\ref{gfgsig}). As soon as for every spin
structure there is the only Green function \cite{dan5}
satisfying eqs.(\ref{trzgr}) , the right side of
(\ref{gfgsig}) is, in fact, independent of $\{\sigma_s\}$.
Furthermore, the $\chi_{F_s}(z')$ conformal 3/2-tensors are
expressed in the terms of $\Phi_{\sigma,N_s}^{(0)}$ in
(\ref{chgsig}) as \cite{dan5}
\begin{equation}
\Phi_{\sigma,F_s}^{(0)}=\sum_{F_r=\mu_r,\nu_r}
\tilde M_{F_s,F_r'}(\{\sigma_q\})\chi_{F_r}
\label{Phipsi}
\end{equation}
where the $\tilde M_{F_s,F_r'}(\{\sigma_q\})$ elements of the
$\tilde M(\sigma)$ matrix are given by
\begin{equation}
\tilde M_{F_s,F_r'}(\{\sigma_q\})=
\int\limits_{C_{v_r}}
\Phi_{\sigma,F_s}^{(0)}(z)\frac{dz} {2\pi i}
P_{F_r'}(z)-\frac{[1-(-1)^{2l_{1_r}}]}{2}
\int\limits_{\tilde C_r}
\Phi_{\sigma,F_s}^{(0)}(z)\frac{dz} {2\pi i}
P_{F_r'}^{(a)}(z).
\label{mtr}
\end{equation}
The desired $\tilde M(\{\sigma_p\})$ matrix in (\ref{zgh}) is
just the same as that defined by (\ref{mtr}).
Eqs.(\ref{Phipsi}) and (\ref{mtr}) are derived from the
condition that, being calculated from (\ref{gfgsig}),
the changes of $G_f(z,z')$ under $2\pi$-twists about
$(A_s,B_s)$-cycles are given by (\ref{trzgr}).
Both the $C_{v_s}$ contours and the $\tilde C_r$ ones are the
same as in (\ref{eqxi}), (\ref{eqy}) and (\ref{gfgsig}).
The $\Phi_{\sigma,F_s}^{(0)}(z)$ conformal 3/2-tensors are
calculated explicitly  from eqs.(\ref{grsig}) and
(\ref{chgsig}). It is useful to note that going certain of
$A_s$-cycles about each other turns both $G_{(\sigma)} (z,z')$
and $\Phi_{\sigma,F_s}^{(0)}(z)$ associated with the $S_{ev}$
superspin structures to the superspin structures containing
pairs of the odd genus-1 superspin ones. Particular, it is
follows from this fact and from (\ref{zgh}) that under the
discussed $S_{ev}\rightarrow S_2$ transformation,
$Z_{S_{ev}}(\{q_{N_s}\})$ holomorphic partition function turns
into $Z_{S_2}(\{q_{N_s}\})$ with exception
only the factor due to both the change of the contours of the
integration in the $\tilde M(\{\sigma_p\})$ matrix (\ref{mtr})
and to the modification of the $(P_{F_1},P_{F_2})$ polynomials
(\ref{poly}) in (\ref{mtr}).  Indeed, before the discussed going
of $C_{u_2}$ about $C_{u_1}$ circle being performed, the
($\tilde C_1,\tilde C_2$) contours in (\ref{mtr}) present none
other than the $(z_s^{(-)})(z_s^{(+)})$ lines in Fig.1(a) where
$z_s^{(+)}=g_s(z_s^{(-)})$ and $s=1,2$. The $z_s^{(-)}$ point is
the intersect of the $C_{v_s}$ circle with the $(v_s)(u_s)$
line. After the going of $C_{u_2}$ about $C_{u_1}$ circle to be
performed, the $(z_s^{(-)})(z_s^{(+)})$ lines are deformed to be
as in Fig.1(b). So the integral along the additional paths
arises. In the $s=2$ case the above integral can be reduced
to the one taken along the $C_{u_1}$ circle together with the
integral along the $(z_1^{(+)})(z_2^{(+)})$ line in Fig.1(c).
In the $s=1$ case the integral along the $C_{u_1}$ circle is
added. To express the discussed $\tilde M(\{\sigma_p\})$ matrix
 in terms of that assigned to the $S_2$ spin structure,
one must also to take into account that the $(P_{F_1},P_{F_2)}$
polynomials (\ref{poly}) associated with the
$S_2$ spin structure are calculated from those associated with
the $S_{ev}$ spin structures by suitable
$\sqrt{k_s}\rightarrow-\sqrt{k_s}$ replacements. Furthermore, the
integral along every $C_{v_s}$ circle can be turned to the
integral along the $C_{u_s}$ circle by the $z\rightarrow
(d_sz-b_s)(-c_sz+a_s)^{-1}$ replacement. In this case one
obtains the expression of $Z_{S_{ev}}(\{q_N\})$ in terms of
$Z_{S_2}(\{\tilde q_N\})$ as follows
\begin{equation}
Z_{S_{ev}}(\{q_N\})=\frac{
Z_{S_2}(\{\tilde q_N\})}{\det(I+\tilde X)}
\label{zrp}
\end{equation}
where the $\tilde X_{F_sF_j}$ elements of the $\tilde X$ matrix
are given by
\begin{eqnarray}
(-1)^j\tilde
X_{F_sF_j'}=\int\limits_{C_{u_1}}\tilde\chi_{F_s}(z')
P_{F_j'}^{(a)}(z')\frac{dz'}{2\pi i}
-2\int\limits_{\tilde
C_{12}}\tilde\chi_{F_s}(z')P_{F_j'}^{(a)}(z')\frac{dz'}{2\pi i}
-\int\limits_{C_{u_2}}\tilde\chi_{F_s}(z')P_{F_j'}^{(a)}(z')
\frac{dz'}{2\pi i}
\nonumber \\
{\rm  for}\quad F_j'=\mu_j,\nu_j \quad{\rm with}\quad j=1,2;
\nonumber \\
\tilde X_{F_sF_j'}=0\quad
{\rm  for}\quad F_j'=\mu_j,\nu_j \quad{\rm with}\quad j>2.
\label{res1}
\end{eqnarray}
In deriving (\ref{zrp}) and (\ref{res1}) we employ
eqs.(\ref{Phipsi}) and (\ref{mtr}). One can see from
(\ref{res1}) and (\ref{res}) that the $\tilde X$ matrix differs
from the $X$ matrix by the $\tilde\chi_{F_s}(z')\rightarrow
\chi_{F_s}(z')$ replacement.  Nevertheless, we prove that
$\det(I+\tilde X)=\det(I+X)$ and, therefore, eq.(\ref{zrp}) is
the same as (\ref{zerap}).

For this purpose we note that the $\tilde\chi_{F_s}(z)$
conformal zero modes in (\ref{res}) are expressed in terms of
the $\chi_{F_s}(z)$ conformal 3/2-tensors as follows
\begin{equation}
\tilde\chi_{F_s}(z)=\chi_{F_s}(z)-\sum_{F_r'\in\{N_0\}}
A_{F_sF_r'} \chi_{F_r'}(z)
\label{chipsi}
\end{equation}
where
$F_p=(\mu_p,\nu_p)$ and the $\{N_0\}$ set of the indices is the
same as in (\ref{ghgr}). The above indices are associated with
those Schottky parameters, which chosen to be the same for all
genus-$n$ supermanifolds. Furthermore, $A_{F_sF_r'}$ elements of
the $A$ matrix are defined by
\begin{eqnarray}
A_{\mu_s\mu_r}=\frac{u_s-v_r}{u_r-v_r},\quad A_{\mu_s\nu_r}=
\frac{u_r-u_s}{u_r-v_r},\quad
A_{\nu_s\mu_r}=\frac{v_s-v_r}{u_r-v_r},\quad
A_{\nu_s\nu_r}=\frac{u_r-v_s}{u_r-v_r}, \nonumber\\
{\rm if}\quad (\mu_r,\nu_r)\in\{N_0\}; \quad
{\rm otherwise}\quad A_{F_sF_r'}=0.
\label{Aff}
\end{eqnarray}
In (\ref{Aff}), like throughout above,  $(u_p,v_p)$ are the
fixed points of the $z\rightarrow g_p(z)$ Schottky
transformation (\ref{sch}).  To derive (\ref{chipsi}) and
(\ref{Aff}), we calculate the $\tilde P_{F_s}$ polynomials in
(\ref{ghgr}) from the condition that in the sum on the right
side of (\ref{chgr}), there are no terms proportional to
$P_{F_s}$ with $F_s\in\{N_0\}$. Employing, in addition,
eqs.(\ref{trzgr}), one obtains both (\ref{chipsi}) and
(\ref{Aff}). Hence the $X$ matrix can be written down as
\begin{equation}
X=\tilde X-A\tilde M^{-1}\tilde X
\label{xtx}
\end{equation}
where the $A$ matrix is defined by (\ref{Aff}). Eq.(\ref{xtx})
follows directly from (\ref{res}), (\ref{Phipsi}), (\ref{res1})
and (\ref{chipsi}).  Furthermore, it is
obvious from (\ref{xtx}) that
\begin{equation}
\det(I+X)=\det(I+\tilde X)\det[1-A\tilde M^{-1}\tilde X
(I+\tilde X)^{-1}]
\label{dtxx1}
\end{equation}
that is the same as follows
\begin{equation}
\det(I+X)=\det(I+\tilde X)\det[1-(I+\tilde X)^{-1}
\tilde XA\tilde M^{-1}]
\label{dtxx2}
\end{equation}
In addition, one can verify by the direct calculation that
\begin{equation}
\tilde XA=0
\label{xa0}
\end{equation}
and, therefore, the desirable relation
\begin{equation}
\det(I+\tilde X)=\det(I+X)
\label{dtxx}
\end{equation}
takes place. So the partition functions (\ref{zgh})
obey eq.(\ref{zerap}), and, therefore, they are covariant under
the particular modular transformations considered.

\section{Supermodular covariance of the superstring partition
functions in the general case}

As it has been noted in the Introduction, partition functions
(\ref{zgh}) have been calculated \cite{dan5} from equations that
are none other than Ward identities. The above equations are
obtained from the condition that the discussed amplitudes are
independent of a choice of {\it vierbein} and the gravitino
field. Hence it is natural to expect that these equations do
be supermodular covariant.  Below we give the direct proof that
the considered equations really possess the covariance under the
supermodular group discussed.  Since the above equations fully
determine the partition functions ( up to  constant factors only
), the partition functions (\ref{zgh}) necessarily satisfy
restrictions due to the whole supermodular group. The desired
equations have the following form \cite{dan1,dan0,dan5}
\begin{equation}
\sum_N \tilde\chi_N(t;L)\frac{\partial}{\partial q_N}\ln\hat
Z_{L,L'}(\{q_N,\overline q_n\})=<T_{gh}+T_m>- \sum_N
\frac{\partial}{\partial q_N} \tilde\chi_N(t;L)
\label{eq}
\end{equation}
together with the equations to be complex conjugated to
(\ref{eq}). The derivatives  with respect to odd moduli in
(\ref{eq}) are implied to be the "right" ones. The
$\tilde\chi_N(t;L)$  superconformal 3/2-tensor zero
modes will be defined below.\footnote{ These superconformal zero
modes are denoted in \cite{dan5} as $\tilde\chi_N(t)$.}
Furthermore, $T_{gh}$ and $T_m$ are the stress tensors of the
ghost and string superfields, respectively. In the explicit form
\begin{equation}
T_m=10(\partial X)D X/2\quad{\rm and}\quad
T_{gh}=-(\partial \hat F)\hat B-\partial(\hat F\hat B)+D[(D\hat
F)\hat B]/2
\label{strsg}
\end{equation}
where $D$ denotes the spinor derivative (\ref{supder}) and $X$
is the scalar superfield, the space-time dimension being 10.
In addition, $\hat B$ is 3/2-tensor ghost superfield and $\hat
F$ is the vector ghost one. In (\ref{strsg}) the explicit
dependence on the supercoordinate $t=(z|\theta)$ is omitted.
The above $T_m$ is calculated in term of the
$G_{(m)}(t,t';L)$ vacuum correlator defined as follows
\begin{equation}
G_{(m)}(t,t';L)=-D(t')\partial_z<X(t,\overline t)
X(t',\overline {t'})>
\label{gmx}
\end{equation}
Furthermore, $T_{gh}$ is calculated in terms of the ghost
superfield vacuum correlator $G_{gh}(t,t',L)$ where
\begin{equation}
G_{gh}(t,t';L)=<\hat F(t,\overline t)\hat
B(t',\overline {t'})>.
\label{fbgr}
\end{equation}
It is quite essential that, unlike the well
known ghost scheme \cite{fried,pol}, the vacuum correlator
(\ref{fbgr}) has depending on $t$ periods  under rounds about
$(A_s,B_s)$-cycles. In the explicit form (the explicit
dependence on $L$ is omitted )
$$G_{gh}(t_s^a,t')=
Q_{\Gamma_{a,s}}^{-2}(t)\left(G_{gh}^{(s)}(t,t')+
\sum_N Y_{a,N}^{(s)}(t)\tilde\chi_N(t')\right),$$
\begin{equation}
G_{gh}(t_s^b,t')=Q_{\Gamma_{b,s}}^{-2}(t)\left(G_{gh}(t,t')+
\sum_N Y_{b,N}^{(s)}(t)\tilde\chi_N(t')\right)
\label{gencase}
\end{equation}
where 3/2-zero modes $\tilde\chi_N$ are  the same as in
(\ref{eq}). Both the $t_s^a=\Gamma_{a,s}(l_{1s})(t)$
transformations and the $t_s^b =\Gamma_{b,s}(l_{2s})(t)$ ones
(\ref{super}) are defined in Section 2.  Furthermore,
$G_{gh}^{(s)}(t,t')$ is obtained from $G_{gh}(t,t')$ by
$2\pi$-twist about $C_{v_s}$-circle (\ref{circ}). At least,
$Y_{a,N}^{(s)}$ and $Y_{b,N}^{(s)}$ are polynomials of degree 2
in $(z,\theta)$. The sum over $N$ in (\ref{gencase}) includes
only those $Y_{p,N_r}^{(s)}(t)$, which associated with the
Schottky parameters that are moduli.
We use for the $\{N_r\}$ indices the same notation
$(k_r,u_r,v_r,\mu_r,\nu_r)$ as for the Schottky parameters.
In this notation, particularly, $q_{k_r}=k_r$,
$q_{u_r}=u_r$ and so one. In this case the above polynomials are
given as follows \cite{dan0,dan5}
\begin{equation}
Y_{p,N_r}^{(s)}(t)=Y_{p,N_s}(t)\delta_{rs}\quad{\rm
where}\quad p=a,b \quad{\rm and}\quad
Y_{p,N_s}(t)=Q_{\Gamma_{p,s}}^2\left[\frac{\partial g_s^p}
{\partial q_{N_s}}+\theta_s^p\frac{\partial \theta_s^p}
{\partial q_{N_s}}\right].
\label{genY}
\end{equation}
Eqs.(\ref{gencase}) being taken into account, the condition for
$G_{gh}(t,t')$ to be 3/2-supertensor on $t'$-supermanifold,
fully determines \cite{dan1,dan0,dan5} both $G_{gh}(t,t')$ and
3/2-superconformal zero modes $\tilde\chi_N(t')$.  At the odd
modular parameters to be zero, eqs.(\ref{gencase}) reduce to
(\ref{chgr}) and (\ref{rgr}). Unlike the ghost correlators
considered in \cite{ver,martnp}, $G_{gh}(t,t')$ has no unphysical
poles \cite{dan1,dan0,dan5}. In the calculation of $T_m+T_{gh}$,
the singularity at $z\rightarrow z'$ in both $G_{(m)}(t,t')$ and
$G_{gh}(t,t')$ is removed by the usual prescription
\cite{fried}.

Eqs.(\ref{eq}) resemble the equations  discussed in
\cite{martnp,fried}.  But, unlike those in \cite{martnp,fried},
they take into account, in addition, the factors due to both
ghost zero modes and the moduli volume form. The above terms are
taken into account in (\ref{eq}) owing to the using of the
$G_{gh}(t,t')$ ghost superfield vacuum correlator satisfying
eqs.(\ref{gencase}) and owing to the presence of the
proportional to $\partial_{q_N}\tilde\chi_N(t)$ terms on the
right side of (\ref{eq}). The difference between eqs.(\ref{eq})
and those in \cite{martnp,fried} is especially urgent for those
superspin structures where at least one of the $l_{1s}$
characteristics is unequal to zero.  Indeed, in this case the
equations \cite{martnp,fried} have no solutions at all.  To the
contrary, (\ref{eq}) allows to obtain \cite{dan5} explicit
formulae for the partition functions associated with the
superspin structures discussed.

One can see that both $T_{gh}$ and $T_m$ have the usual form
\cite{fried} in the terms of the ghost or string superfields,
but in the considered scheme, $T_m+T_{gh}$ is not  superconformal
3/2-form under the $(\Gamma_{a,s}(l_{1s}),\Gamma_{b,s}(l_{2s}))$
mappings because the $G_{gh}(t,t')$ ghost superfield vacuum
correlators have the periods under the mappings above.
Nevertheless, the right side of eq.(\ref{eq}), as well as the
left side, appears to be superconformal 3/2-form under the
considered mappings.  Moreover, we prove that
eqs.(\ref{gencase})  are covariant under supermodular
transformations.

As it was already noted in this paper, the discussed
transformations, generally, present globally defined
$t\rightarrow\hat t(t,\{q_N\})=(\hat z|\hat\theta)$ mappings
that accompanied by both the $L\rightarrow \hat L$ change of the
superspin structure and the change $q_N\rightarrow \hat
q_N(\{q_M\})$ of the moduli. Under the considered
transformations, $G_{gh}(t,t';L)\rightarrow
G_{gh}(\hat t,\hat t';\hat L)$ and  $G_{(m)}(t,t';L)\rightarrow
G_{(m)}(\hat t,\hat t';\hat L)$. Since  $G_{(m)}(t,t';L)$ is
defined by (\ref{gmx}) to be the  tensor under
globally defined superconformal transformations, the desired
$G_{(m)}(\hat t,\hat t';\hat L)$ is given by
\begin{equation}
G_{(m)}(t,t')=\hat Q(\hat
t)G_{(m)}(\hat t,\hat t')\hat Q^2(\hat t)
\label{trgm}
\end{equation}
To calculate $G_{(m)}(\hat t,\hat t';\hat L)$, it is useful to
note that a number of rounds about $(A,B)$ cycles on $\hat t$
supermanifold corresponds to every $2\pi$-twist about either
$A$-cycle or $B_s$-cycle on the $t$ one. Therefore, to every
$t\rightarrow \Gamma_{b,s}(l_{2s})(t)$ mapping and to every
$t\rightarrow \Gamma_{a,s}(l_{1s})(t)$ mapping, the appropriate
mappings on $\hat t$-supermanifold can be assigned. For the
condition that under the above mappings, $G_{(m)}(\hat t,\hat
t';\hat L)$ is changed in the accordance with
eqs.(\ref{gencase}) written in terms of the variables assigned
to $\hat t$-supermanifold, the desired supermodular
transformation of $G_{gh}(t,t';L)$ turns out to be
\begin{equation}
G_{gh}(t,t';L)=\hat Q^{-2}(\hat t)\left(G_{gh}(\hat t,\hat
t';\hat L)+ \sum_N\left[\frac{\partial z(\hat t)}{\partial \hat
q_N}
+\theta(\hat t)\frac{\partial\theta(\hat t)}{\partial \hat
q_N}\right]\tilde\chi_N(\hat t';\hat L)\right)\hat Q^3(\hat t').
\label{smodgr}
\end{equation}
In (\ref{smodgr}) the $\hat Q(t)$ factor is defined to be
\begin{equation}
\hat Q^{-1}(\hat t)=D(\hat t)\theta(\hat t')
\label{hatq}
\end{equation}
and $D(\hat t)$ is the spinor derivative  (\ref{supder}) with
respect to $\hat t=(\hat z|\hat\theta)$. Furthermore, the
$\tilde\chi_N(t,L)$ superconformal 3/2-zero modes in
(\ref{gencase}) are written down in terms of $\chi_N(\hat
t';\hat L)$ in (\ref{smodgr}) as follows
\begin{equation}
\tilde\chi_N(t,L)=\sum_{N'}\frac{\partial q_N}{\partial\hat q_N'}
\tilde\chi_{N'}(\hat t;\hat L)\hat Q^3(\hat t)
\label{hatchi}
\end{equation}
In addition to eqs.(\ref{smodgr}), (\ref{hatchi}) and
(\ref{trgm}), one must take into account eq.(\ref{sdet}), which
describes the supermodular transformation of the partition
functions. In this case one can verifies by the direct
calculation that eqs.(\ref{eq}) appear to be covariant under the
transformation discussed.

\section{ Supermodular invariance of the multi-loop superstring
amplitudes}

In the self-consistent theory the multi-loop
superstring amplitudes $A_n$ must satisfy the restrictions due
to the supermodular group.  Naively, eq.(\ref{ampl}) satisfies
the above restrictions because every even ( odd ) superspin
structure contribution in $A_n$ can be derived by
supermodular transformations of the contribution due to a fixed
even ( odd ) structure. In fact, however, the supermodular
invariance of eq.(\ref{ampl}) must be ensured by a suitable
regularization procedure because the integration of every
superspin contribution in (\ref{ampl}) is divergent. It is
worth-while to note that the above regularization procedure is
necessary even if in the whole integrand the singularities
are cancelled after the summation over the superspin
structures to be performed.
Indeed, the
$(q_N\rightarrow\hat q_N, t^{(r)}\rightarrow\hat t^{(r)})$
change, being associated with the particular supermodular
transformation, depends on the superspin structure in terms
proportional to odd modular parameters, as it has been shown in
the previous Sections. As the result, the integrand in
(\ref{ampl}) appears to be non-covariant under the discussed
transformations. In this case the supermodular invariance of
(\ref{ampl}) could be the result of an appropriate integration
of every superspin structure contribution. Being divergent, this
integration needs the regularization procedure. To avoid the
above regularization procedure, it seems
attractive to re-write down the right side of (\ref{ampl}) to be
the integral of the supermodular covariant function. For this
purpose we assign to every superspin structure contribution in
(\ref{ampl}) the suitable mapping
$[t\rightarrow t_L(t;\{q_N\}), q_N\rightarrow
q_{LN}(\{q_N\})]$.  Furthermore, we write down the desired $A_n$
amplitude as follows
\begin{equation}
A_n=\int\prod_N dq_Nd\overline
q_N\prod_r dt^{(r)}d\overline t^{(r)} I(\{q_N,\overline
q_N\};\{t^{(r)},\bar t^{(r)}\}).
\label{ampl1}
\end{equation}
In (\ref{ampl1}), the $I(\{q_N,\overline q_N\};\{t^{(r)},\bar
t^{(r)}\})$ integrand is defined as
\begin{eqnarray}
I(\{q_N,\overline q_N\};\{t^{(r)},\bar t^{(r)}\})=
\sum_{L,L'}\hat Z_{L,L'}(\{q_{LN}\},\{\overline q_{L'N}\})
Jac(\partial q_{LN}/\partial q_{N'}) Jac(\partial \overline
q_{L'N}/\partial \overline q_{N'})
\nonumber \\
\times\left(\prod_r Jac(\partial t_{L}^{(r)}/\partial
t^{(r)}) Jac(\partial \overline t_{L'}^{(r)}/\partial \overline
t^{(r)})\right) <\prod_rV(t_L^{(r)},\overline t_{L'}^{(r)})>
\label{int}
\end{eqnarray}
where $Jac(\partial q_{LN}/\partial q_{N'})$ and
$Jac(\partial t_L^{(r)}/\partial t^{(r)})$ are the Jacobians of
the corresponding transformations. It is implied that the
consideration given in the previous Sections is referred to
$t_L(t,\{q_N\})$ and $q_{LN}(\{q_N\})$. Particular, eqs.
(\ref{smodtr}), (\ref{eqxi})-(\ref{eqych}), (\ref{eqodd}),
(\ref{hatxi}) and (\ref{munu}) determine the action of
the supermodular group on $t_L$ and $q_{LN}$. The $(t,\{q_N\})$
dependence of $t_L$ and of $q_{LN}$ is calculated from
the condition that  the $(t\rightarrow\hat t,q_N\rightarrow\hat
q_N)$ change under every supermodular transformation
associated with the given integral matrices in (\ref{modtr}) and
(\ref{smodtr}) is the same for all the superspin
structures. In this case the integrand in (\ref{ampl1})
appears covariant under the supermodular transformations.
The region of the integration over even moduli $q_{ev}$
in (\ref{ampl1}) is the quotient of the $q_N$ space by the
supermodular group.

Without loss of generality, one can take $t=t_{L_0}$ and
$q_N=q_{L_0N}$ for the $L_0$ superspin structure. We
choose $L_0$ to be the superspin structure $S(0)$  where
$l_{1s}=l_{2s}=0$ for every $s$. It is convenient because
the supermodular transformations discussed in Sections 3 and 4
map the $S(0)$ superspin structure onto itself. In this case
action of the supermodular group on $t$ and $q_N$ is
determined by (\ref{smodtr}), (\ref{eqxi})-(\ref{eqych}),
(\ref{eqodd}) and (\ref{hatxi}), all they being taken for
$L=S(0)$. Particular, to calculate the quadratic in
$\{\mu_s,\nu_s\}$ terms in $\hat q_{ev}$, one can substitute in
(\ref{smodtr}) eqs. (\ref{eqodd}) with $\eta^{(qr)}=0$ and
with the $\hat X$ matrix taken at zero odd super-Schottky
parameters. After the above substitution to be performed in
(\ref{smodtr}), eqs. (\ref{smodtr}) determine in the discussed
approximation the integration region in (\ref{ampl1}).

Under the above $L_0=S(0)$ choice, the $(t=t_L,q_N=q_{LN})$
relations take place for all the Neveu-Schwarz superspin
structures $S_1$ ( in this case, $l_{1s}=0$ for
every $s$ ).  Indeed, as it was discussed in
Section 2, all these superspin structures can be
derived from $S(0)$ by the $\sqrt {k_s}\rightarrow-\sqrt{k_s}$
replacements.  Furthermore, for the superspin structures with
non-zero $l_{1s}$-characteristics, the $(t_L,q_{LN})$ variables
differ from $(t,q_N)$ only by terms proportional to the odd
modular parameters. Employing the results obtained in the
previous Sections, one can calculate $t_L(t;\{q_N\})$ and
$q_{LN}(\{q_N\})$ assigned to the above superspin structures.
For superspin structures $S_{ev}$  without the
odd genus-1 superspin ones we use the supermodular
transformations discussed in Section 3. In this case the desired
relations for the calculation of $t_L(t;\{q_N\})$ and
$q_{LN}(\{q_N\})$ are given by
\begin{equation}
\hat t_{\hat L}(t_L,\{q_{LN}\})=t_{\hat L}(\hat t_0,\{\hat
q_{N0}\})\,.
\label{www}
\end{equation}
On the left side of (\ref{www}), $t_L\equiv
t_L(t,\{q_N\})$ and $q_{LN}\equiv q_{LN}(\{q_N\})$. Furthermore,
$\hat t_0\equiv \hat t_0(t,\{q_N\})$ and $\hat q_{N0}\equiv\hat
q_{N0}(\hat q_N)$. In (\ref{www}) the  $(t\rightarrow\hat
t_0,q_N\rightarrow q_{N0})$ supermodular transformations are
calculated for the $S(0)$ superspin structure defined above.
Eqs.(\ref{www}) follow directly from the condition that the
right side of (\ref{int}) is covariant under the
supermodular transformations considered. In the $L=S(0)$ case
eqs.(\ref{www}) degenerate to be the identity.  To solve
(\ref{www}) for the $L\neq S(0)$, one can take $\hat L=S_1$. In
this case $t_{\hat L}(t)=t$. Hence eqs. (\ref{www}) determine
both $t_L(t,\{q_N\})$ and $q_{LN}(\{q_N\})$ for all the $S_{ev}$
superspin structures discussed. Particular, in the linear
approximation in $\{\mu_r,\nu_r\}$, the desired
$\{\mu_{Lr},\nu_{Lr}\}$ for $L=S_{ev}$  are given by
\begin{eqnarray}
\mu_{Lr}=\sum_{s=1}^n\left[\left(\hat X(L)\hat
X^{-1}(L_0)\right)_{\mu_r\mu_s}\mu_s+ \left(\hat X(L)\hat
X^{-1}(L_0)\right)_{\mu_r\nu_s}\nu_s\right]\,,
\nonumber\\
\nu_{Lr}=\sum_{s=1}^n\left[\left(\hat X(L)\hat
X^{-1}(L_0)\right)_{\nu_r\mu_s}\mu_s+ \left(\hat X(L)\hat
X^{-1}(L_0)\right)_{\nu_r\nu_s}\nu_s\right]
\label{eqodl}
\end{eqnarray}
where both $\hat X(L)$ and $\hat X(L_0)$ are taken at zero odd
super-Schottky parameters and $L_0=S(0)$. Eqs. (\ref{eqodl})
follows from (\ref{eqodd}) and (\ref{www}). Furthermore, if one
substitute (\ref{eqodl}) into (\ref{smodtr}), one can calculate
from the obtained equations the region of the integration  over
$\{q_{ev}\}$ in the quadratic approximation in
$\{\mu|s,\nu_s\}$. The above integration region turns out to be
calculated in terms of the modular group parameters at zero odd
moduli and in terms of the odd super-Schottky parameters, as
well.To calculate the $t_L(t,\{q_N\})$ functions for the even
superspin structures containing the odd genus-1 superspin ones,
one must consider the transformations discussed in Section 4. In
this case (\ref{www}) is replaced as
\begin{equation}
\hat t_{\hat
L}(t_L,\{q_{LN}\})=t_{\hat L}^{(-)}(t,\{q_N\})
\label{www2}
\end{equation}
where $t_L\equiv t_L(t,\{q_N\}),q_{LN}\equiv q_{LN}(\{q_N\})$
and $t_{\hat L}^{(-)}(t,\{q_N\})$ is calculated from $t_{\hat
L}(t,\{q_N\})$ by the going of suitable $A_s$-cycles about each
other.  Starting with $\hat L$ to be among the $S_{ev}$
superspin structures, one calculate from (\ref{www2}) both
$t_L(t,\{q_N\})$ and $q_{LN}(\{q_N\})$ for the superspin
structures containing a number of pairs of the odd genus-1
superspin ones. This calculation is quite similar to that
performed for the $S_{ev}$ superspin structures.

It is obvious that (\ref{ampl1}) should be reduced to
(\ref{ampl}), if the integration of every superspin contribution
were to be finite.  The above integrations being divergent, we
define the $A_n$ amplitude by eq.(\ref{ampl1}). Unlike
(\ref{ampl}), eq.(\ref{ampl1}) satisfies in the explicit form
all the restrictions due to the supermodular group. Hence the
study of the divergency problem in the considered theory is
reduced to the investigation of the singularities of (\ref{int}).
Particular, the absence of non-integrable singularities in
(\ref{int}) means the finiteness of the theory discussed.

Generally, it is expected \cite{martnp,martpl,mandel,berk,davis}
that the (super)modular invariance provides the finiteness of
the superstring theory. The reason is \cite{martnp,mandel} that
the above invariance origins the space-time supersymmetry of the
Ramond-Neveu-Schwarz superstrings. In turn, the space-time
supersymmetry prohibits the tadpoles appearing to be the only
source of possible divergencies in the theory. So one can hope
that being supermodular covariant, (\ref{int}) is free from
non-integrable singularities. The study of potential
singularities of (\ref{int}) and, therefore, potential
divergencies of (\ref{ampl1}) requires the detailed
investigation of the modular measure \cite{dan5} that goes out
of the framework of this paper. In the present paper we restrict
ourselves only by a brief consideration of the subject
discussed.

One can see from eq.(\ref{hol}) together with eqs. (\ref{znsch}),
(\ref{hhh}), (\ref{b3}), (\ref{z0m}) and (\ref{z0gh}) of Appendix
B that singularities may arise in (\ref{int}), if
bodies of the $k_{Ls}(\{q_N\})$ multipliers assigned to
basic Schottky transformations go to whether unity ( up to the
phase ) or zero.  It is follows from (\ref{znsch}), (\ref{b3}),
(\ref{z0m}) and (\ref{z0gh}) that singularities  present in
(\ref{int}) also when bodies of multipliers $k$ assigned to
products of the basic Schottky transformations go to unity ( up
to the phase ). One expects, however, that the last
singularities do not origin a divergence of (\ref{ampl1})
because the $k\rightarrow1$ limit does not mean the
degenerateness of the Riemann surface. At the same time,
potential divergences in string theories arise from the
degenerateness of Riemann surfaces \cite{belkniz}. In fact the
discussed singularity in (\ref{int}) is compensated by a
smallness of the integration volume associated with the
configurations considered. Moreover, the domain where bodies of
the $k_{Ls}(\{q_N\})$ Schottky multipliers are near to unity ( up
to the phase ) is equivalent modulo of modular group to the
domain where bodies of these Schottky multipliers are small
\cite{martpl}. So the above domain can be excluded from the
integration region. Furthermore, one can see from (\ref{eqych}),
(\ref{hdel}) and (\ref{poly}) that vanishing the
$k_{Ls}(\{q_N\})$ body appears when $k_s\rightarrow0$. In this
case $k_{Ls}(\{q_N\})\rightarrow k_s[1+o_s(L)]$ where $o_s(L)$
is proportional to odd modular parameters. The highest
$k_s^{-3/2}$ singularity appears in the case when
$(l_{1s}=0,l_{2s}=1/2)$ or $(l_{1s}=l_{2s}=0)$. As it is usual
\cite{vec8}, in every sum of two superspin structures
distinguished only in the discussed genus-1 superspin ones, the
above $k_s^{-3/2}$ singularity is reduced to $k_s^{-1}$ because
$o_s(L)$ is the same for both the superspin structures
considered. In addition, the non-holomorphic factor in
(\ref{hol}) gives the factor $(\ln|k|)^{-5}$ in (\ref{int}). As
the result, the integration over small $k_s$ does not lead to
divergency of (\ref{ampl1}).  Moreover, one finds to be
finite the integral over the region where the fixed points
associated with a particular basic Schottky transformation go
away from each other. Indeed, in this case the radius of the
circles (\ref{circ}) associated with the considered Schottky
transformation is taken to be finite.  Otherwise the above
circles intersect the ones associated with other basic Schottky
transformations. The finiteness of the above radius at
$|u_s-v_s|\rightarrow\infty$ requires $k_s$ to be small as
$|k_s|\equiv|u_s-v_s|^{-1}$ that provides the finiteness of the
integral discussed.

Because of the $(u_s-v_s-\mu_s\nu_s)^{-1}$ factors in
(\ref{hhh}), a potential singularity in (\ref{int}) is also
expected if, for a particular handle, $|u_s-v_s|\rightarrow0$.
The above singularity could lead to  divergencies of
(\ref{ampl}). In the considered $|u_s-v_s|\rightarrow0$ limit
the genus-$n$ Riemann surface is degenerated in two separate
Riemann, one of genus 1 and the other genus $(n-1)$. If a number
of the vertices in (\ref{int}) present on both the above
surfaces, the discussed singularity origins the threshold
singularities of (\ref{ampl}) at suitable magnitudes of the
external 10-momenta.  But in the configuration where all the
vertices appear to be whether on the genus-1 surface or on the
one of genus $(n-1)$, the considered singularity should cause a
divergency of (\ref{ampl}) independent of 10-momenta above. One
can see from (\ref{int}) and (\ref{hhh}) that in the discussed
region the integrand of (\ref{ampl1}) has the following form
\begin{equation}
\left|\frac{1}{u_s-v_s}
+\frac{\mu_s\nu_s+\hat o_s(L)}{(u_s-v_s)^2}
\right|^2\left[I_1^{(0)}I_{n-1}+I_1I_{n-1}^{(0)}+(u_s-v_s)B+
(\bar u_s-\bar v_s)\bar B\right]
\label{final}
\end{equation}
where $I_m$ is the integrand for the genus-$m$ amplitude
( with $m=n-1$ or $m=1$ ) and $I_m^{(0)}$ is the same for the
genus-$m$ vacuum one. For $m=1$, the discussed integrand is
obtained by the factorization of (\ref{int}) when the
particular handle moves away from the others.  Furthermore,
$\hat o_s(L)$ appears in (\ref{final}) because of the difference
between $(u_s,v_s)$ and the fixed points of the Schottky
transformations. The line over denotes the complex
conjugation and $B$ describes  the terms proportional to
$(u_s-v_s)$. One can see, for the discussed singularity
to be absent in (\ref{final}), the necessary condition is
$I_m^{(0)}=0$. The above $I_m^{(0)}=0$ could be the consequence
of the space-time supersymmetry, which causes the vanishing of
the vacuum amplitude \cite{martpl}. Using the measure \cite{dan5}
presented in Appendix B, one can show without essential
difficulties that $I_1^{(0)}=0$, but the verification of the
discussed statement for $m\geq1$ needs an additional
investigation. Furthermore, being necessary, the above
$I_m^{(0)}=0$ condition is insufficient to remove the
singularity in question because the second order pole presents
in (\ref{final}) due to the expansion in the series over the odd
super-Schottky parameters.  So for the discussed singularity to
be absent, the $B=0$ condition must be added. One can see a
reason for this $B=0$ condition to be because the second order
pole in (\ref{final}) is reduced to the first order one
\cite{berk,davis} by a choice of the appropriate variables
\cite{pst}. It is not evidently, however, whether the above
choice is consistent with the supermodular invariance. So an
additional study of the discussed subject seems to be necessary.
The kindred divergencies appear in (\ref{ampl1}) when the
Riemann surface is degenerated in two separate Riemann surfaces,
one of genus $n_1$ and the other of genus $(n-n_1)$ with $n_1>$
and $n-n_1>1$. The integral over the vertex local coordinates
is potentially divergent, too. Indeed, when all the
vertices move to be closed together, the vacuum expectations of
the vertex product in (\ref{int}) are ceased to be independent
of $q_N$.  The discussed vacuum expectations begin to be
covariant under the superconformal extension of the $SL_2$ group
that originates the divergency of the integral over the vertex
coordinates. In this case the  singularity in the integrand is
appear to be similar to (\ref{final}). To remove the above
singularity,the vanishing of the vacuum amplitude is again
necessary,  but insufficient. For the similar reasons,
divergency might arise from the region where all the vertices
move away from each other.

It is necessary to note that we uniquely calculate  the
supercovariant integrand (\ref{int}) taking into account only a
part of the supermodular transformations. So, to be sure in the
self-consistency of the discussed scheme, one should verify that
the above integrand is covariant under the whole supermodular
group. This verification requires, however, a more detailed study
of the discussed supermodular transformations that is not
finished at present. We plan to discuss this problem
in  another paper.  \\ \\

{\bf Acknowledgments} \\ \\
The research described in this publication was made possible
in part by Grants No. NO8000 and No. NO8300 from the
International Science Foundation and in part by Grant No.
93-02-3147 from the Russian Fundamental Research Foundation.

\appendix
\section{}
To give the explicit definitions of the $\rho_s^{(pq)}$
functions in (\ref{main1}) and the $\eta_s^{(pq)}$ functions in
(\ref{main2}) for $p=a,b$ and $q=a,b$ we present the above
functions as
\begin{eqnarray} \rho_s^{(bb)}(z)= \frac{\hat
c_s^{(0)}f_0'(z)y^2(z)}{\hat c_s^{(0)}f+ \hat
d_s^{(0)}}+\frac{\hat\rho_s^{(b)}(z,l_{2s})}
{f_0'(g_s)g_s'(z)}+\rho_s^{(b)}(z),\quad
\rho_s^{(ab)}(z)=
\frac{\hat\rho_s^{(a)}(z,l_{2s})}
{f_0'(g_s)g_s'(z)}+\rho_s^{(b)}(z),\nonumber\\
\rho_s^{(ba)}(z)=
\frac{\hat c_s^{(0)}f_0'(z)y^2(z)}{\hat c_s^{(0)}f+
\hat d_s^{(0)}}+\frac{\hat\rho_s^{(b)}(z,l_{1s})}
{\partial_z f_0^{(s)}(z)}+\rho_s^{(a)}(z),\quad
\rho_s^{(aa)}(z)=
\frac{\hat\rho_s^{(a)}(z,l_{1s})}
{\partial_z f_0^{(s)}(z)}+\rho_s^{(a)}(z),\nonumber\\
\eta_s^{(bb)}(z)=\hat\eta_s^{(b)}(z,l_{2s})+\eta_s^{(b)}(z),
\quad\eta_s^{(ab)}(z)=\hat\eta_s^{(a)}(z,l_{2s})+\eta_s^{(b)}(z),
\nonumber\\
\quad\eta_s^{(ba)}(z)=\hat\eta_s^{(b)}(z,l_{1s})+\eta_s^{(b)}(z),
\quad\eta_s^{(aa)}(z)=\hat\eta_s^{(a)}(z,l_{1s})+\eta_s^{(a)}(z)
\label{def}
\end{eqnarray}
where both $\rho_s^{(p)}(z)$ and $\eta_s^{(p)}(z)$ with
$p=a,b$ are defined as follows
\begin{eqnarray}
\eta_s^{(b)}(z)=\epsilon_s(z,l_{2s})\xi(g_s)\xi'(g_s)+
\xi'(g_s)(z-g_s)\varepsilon_s \varepsilon_s'
+(-1)^{2l_{2s}}\xi(g_s)\varepsilon_s\varepsilon_s'
[c_sz+d_s+(-1)^{2l_{2s}}], \nonumber \\
\eta_s^{(a)}(z)=
[1-(-1)^{2l_{1s}}][\xi^{(s)}(z)\varepsilon_s\varepsilon_s'
-\varepsilon_s(z)\xi^{(s)}\xi^{(s)'}(z)],
\nonumber\\
\rho_s^{(b)}(z)= -(-1)^{2l_{2s}}
\frac{[\varepsilon_s(z,l_{2s})
\varepsilon_s'(z,l_{2s})f'(z)[z-g_s(z)]+
f'(g_s)\epsilon_s(z,l_{2s})\xi(g_s)]}
{f_0'(g_s)g_s'(z)[c_sz+d_s]}, \nonumber\\
\rho_s^{(a)}(z)=[1-(-1)^{2l_{1s}}]
\frac{\varepsilon_s(z)\xi^{(s)}(z)
f^{(s)'}(z)}
{f_0^{(s)'}(z)}.\,\,
\label{rhoeta}
\end{eqnarray}
Furthermore, both $\hat\rho_s^{(p)}(z,l_s)$ and
$\hat\eta_s^{(p)}(z,l_s)$ with $p=a,b$ in (\ref{def}) are given
by
\begin{eqnarray}
\hat\rho_s^{(b)}(z,l_s)=(-1)^{2l_s}
\frac{\hat\varepsilon_s\hat\varepsilon_s'[f-\hat
g_s(f)]}{\hat c_sf+\hat d_s}
-\frac{\xi(z)\hat\epsilon_s(f,l_s)\sqrt{f'(z)}}
{[\hat c_sf+\hat d_s]^2},\nonumber\\
\hat\rho_s^{(a)}(z,l_s)=[1-(-1)^{2l_s}]\xi(z)
\hat\varepsilon_s(f)\sqrt{f'(z)},
\nonumber\\
\hat\eta_s^{(a)}(z,l_s)=[1-(-1)^{2l_s}]
\left[\xi(z)\hat\varepsilon_s\hat\varepsilon_s'-
\frac{\xi(z)\xi'(z) \hat\varepsilon_s(f)}{2\sqrt{f'(z)}}\right]
\nonumber\\
\hat\eta_s^{(b)}(z,l_s)=\frac{\xi(z)\xi'(z)\hat
\epsilon_s(f,l_s)}
{2\sqrt{f'(z)}}- \xi(z)\hat\epsilon_s(f,l_s)
\hat\varepsilon_s'(f)
\label{hrhoeta}
\end{eqnarray}
where $f\equiv f(z)$.

\section{Measure in terms of super-Schottky parameters}
For the Neveu-Schwarz sector matrix elements
$\omega_{ps}(\{q_{N_s}\},L)$ of the period matrix in (\ref{hol})
are given by \cite{vec8,pst}
\begin{eqnarray}
2\pi i \omega_{ps}(\{q_{N_s}\},L)=
{\sum_\Gamma}^{''}\ln\frac{[u_s-g_\Gamma(u_p,\mu_p)-
\mu_s\theta_\Gamma(\mu_p,u_p][v_s-g_\Gamma(v_p,\nu_p)-
\nu_s\theta_\Gamma(\nu_p,v_p]}{[u_s-g_\Gamma(v_p,\nu_p)-
\mu_s\theta_\Gamma(\nu_p,v_p][v_s-g_\Gamma(u_p,\mu_p)-
\nu_s\theta_\Gamma(\mu_p,u_p]}\nonumber\\
+\delta_{ps}\ln k_s\,.
\label{b1}
\end{eqnarray}
In (\ref{b1}) the summation is performed over all
super-Schottky group transformations
$\Gamma=\{z\rightarrow
g_\Gamma(z,\theta),\theta\rightarrow \theta_\Gamma(\theta,z)\}$
except those that have the leftmost to be a power of $\Gamma_s$,
the rightmost being a power $\Gamma_p$.  Besides, $\Gamma\neq
I$, if $s=p$. The $Z_L(\{q_{N_s}\})$ factor in (\ref{hol}) for
the Neveu-Schwarz sector has been found to be
\cite{vec8,pst,dan1,dan0}
\begin{eqnarray}
Z_L(\{q_{N_s}\})=H(\{q_{N_s}\})
\left(\prod_{s=1}^n\frac{(-1)^{2l_{2s}-1}
[1-(-1)^{2l_{2s}}\sqrt{k_s}]^2}{k_s^{3/2}}\right)
\prod_{(k)}[1-(-1)^{\Sigma_k}\sqrt k]^{-2}\nonumber\\
\times\prod_{m=1}^\infty
\frac{[1-(-1)^{\Sigma_k}k^{m-1/2}]^{8}}{(1-k^m)^{8}}
\label{znsch}
\end{eqnarray}
where the product over $(k)$ is taken over all
the multipliers of super-Schottky group (\ref{super}), which are
not powers of other the ones and
\begin{equation}
\Sigma_k=\sum_r(2l_{2r}-1)n_r(\Gamma)\,.
\label{sinsch}
\end{equation}
In (\ref{sinsch}), $n_r(\Gamma)$ is the
number of times that the $\Gamma_r$ generators  are present
in $\Gamma$ (for its inverse $n_r(\Gamma)$ is defined to be
negative ). At last, $H(\{q_{N_s}\})$ in
(\ref{znsch}) is defined by
\begin{equation}
H(\{q_{N_s}\})=g^{2n}(u_1-u_2)(v_1-u_2)
\left[1-\frac{\mu_1\mu_2}{2(u_1-u_2)}-
\frac{\nu_1\mu_2}{2(v_1-u_2)}\right]
\prod_{s=1}^n(u_s-v_s-\mu_s\nu_s)^{-1}
\label{hhh}
\end{equation}
where $g$ is the coupling constant. It is assumed
that $u_1,v_1,u_2,\mu_1$ and $\nu_1$  are fixed to be the same
for all the genus-$n$ supermanifolds and, therefore, they are
not the moduli.

The Ramond sector to be considered, we present
$Z_L(\{q_{N_s}\})$ in (\ref{hol}) for every even superspin
structure as follows \cite{dan5}
\begin{eqnarray}
Z_L(\{q_{N_s}\})= Z_{0(m)}(\{k_s,u_s,v_s\},L)
Z_{0(gh)}(\{k_s,u_s,v_s\},L)\times \nonumber \\
H(\{q_{N_s}\})
\Upsilon_m^{(n)}(\{q_{N_s}\},L)
\Upsilon_{gh}^{(n)}(\{q_{N_s}\},L)
\label{b3}
\end{eqnarray}
where $H(\{q_{N_s}\})$ is given by (\ref{hhh}) and the subscript
"gh" ( "m" ) labels the ghost ( respectively,
string superfield ) contributions. Both
$Z_{0(m)}(\{k_s,u_s,v_s\},L)$
and $Z_{0(gh)}(\{k_s,u_s,v_s\},L)$ are calculated at zero  odd
Schottky parameters. Both
$\ln\Upsilon_m^{(n)}(\{q_{N_s}\},L)$ and
$\ln\Upsilon_{gh}^{(n)}(\{q_{N_s}\},L)$  present the terms
proportional to the odd Schottky parameters. The
$Z_{0(m)}(\{k_s,u_s,v_s\},L)$ factor in (\ref{b3}) is given by
\begin{equation}
Z_{0(m)}(\{k_s,u_s,v_s\},L)=
\frac{\Theta^5[l_1,l_2](0|\omega^{(r)})}
{\Theta^5[\{0\},\{1/2\}] (0|\omega^{(r)})}\prod_{(k)}
\prod_{m=1}^\infty \frac{(1-k^{m-1/2})^{10}}{(1-k^m)^{10}}
\label{z0m}
\end{equation}
where $\Theta$ is the theta function. The $\Theta$ in the
denominator associates with that  spin structure where for every
handle, $l_{1s}=0,l_{2s}=1/2$ . The period matrix $\omega^{(r)}$
is given by (\ref{b1}) taken at zero odd Schottky parameters.
The product over $(k)$ is taken over all
the multipliers of the Schottky group (\ref{sch}), which are
not powers of other the ones. The
$Z_{0(gh)}(\{k_s,u_s,v_s\},L)$ factor in (\ref{b3})
is given by
\begin{eqnarray}
Z_{0(gh)}(\{k_s,u_s,v_s\},L)=\frac{\exp[-\pi i
\sum_{j,r}l_{1j}l_{1r}\omega_{jr}^{(r)}]}
{\sqrt{\det\tilde M(\{\sigma_p\})
\det\tilde M(\{-\sigma_p\})}}
\left[\prod_{s=1}^n\tilde Z_0(k_s;l_{1s},l_{2s})\right]
\nonumber\\
\times\prod_{(k)}\prod_{m=1}^\infty\frac{(1-k^{m+1})^2}
{[1-\Lambda(k,\{\sigma_p\})k^{m+1/2}]
[1-\Lambda(k,\{-\sigma_p\})k^{m+1/2}]}
\label{z0gh}
\end{eqnarray}
where, as in (\ref{z0m}), the product over $(k)$ is taken over
all the multipliers of the Schottky group (17), which are not
powers of the other ones. The period matrix $\omega^{(r)}$
is given by (\ref{b1}) at zero odd Schottky parameters. The
$\tilde M(\{\sigma_p\})$ matrix is defined by (\ref{mtr}).
Furthermore,
\begin{equation}
\Lambda(k,\{\sigma_p\})=\exp\Omega_{\Gamma_{(k)}}(\{\sigma_p\})
\end{equation}
where $\Omega_{\Gamma_{(k)}}(\{\sigma_p\})$ is given by
(\ref{omgm}) for the group products of the basic Schottky
transformations having the multiplier to be $k$.  The $\tilde
Z_0(k_s;l_{1s},l_{2s})$ factors in (\ref{z0gh}) are defined by
\begin{equation}
\tilde Z_0(k_s;l_{1s},l_{2s})=
\frac{(-1)^{2l_{1s}+2l_{2s}-1}
[1-(-1)^{2l_{2s}}\sqrt{k_s}]^2}{4^{2l_{1s}}k_s^{3/2}}\,.
\label{z0s}
\end{equation}
Eq.(\ref{z0s}) is slightly different from eq.(147) in
\cite{dan5} because we use in (\ref{z0gh}) the  $\tilde
M(\{\sigma_p\})$ matrix instead of $M{(0)}(\{\sigma_p\})$
employed in \cite{dan5}. It is useful to remind that in
(\ref{z0m}) and in (\ref{z0gh}), the $k$ multipliers are
calculated at zero odd Schottky parameters. Both
$\Upsilon_m^{(n)}$ and $\Upsilon_{gh}^{(n)}$ in (\ref{b3})
depending on the above odd parameters have the following form
\begin{eqnarray}
\ln\Upsilon_{gh}^{(n)}(\{q_{N_s}\},L)=
trace\ln\left[I+\tilde\Delta_{gh}(\{\sigma_p\})\right]
-\ln\det\hat U(\{\sigma_p\}]
\nonumber\\
+\ln
sdet[U(\{\sigma_p\})U_0^{-1}(\{\sigma_p\})]
\,,
\label{upsil}
\end{eqnarray}
\begin{equation}
\ln\Upsilon_m^{(n)}(\{q_{N_s}\},L)=
-5trace\ln\left[1+\tilde\Delta_m\right]
\label{upsilm}
\end{equation}
where $\tilde\Delta_m$ and $\tilde\Delta_{gh}$ are
integral operators and both $U(\{\sigma_p\})$ and  $\hat
U(\{\sigma_p\})$ to be matrices, all they being defined
below. Furthermore, $U_0(\{\sigma_p\})$ is $U(\{\sigma_p\})$ at
zero odd Schottky parameters.\footnote{Eq.(\ref{upsilm})
corresponds to eq.(134) of \cite{dan5}. In (\ref{upsilm}) we
retrieved an factor $-5$ and symbol "{\it trace}" missed
mistakenly in front of the right side of above eq.(134). In
addition, in \cite{dan5} a number of other inaccuracies sliced
in formulas for the factors considered.  In discussed eq.(134)
of \cite{dan5} the expression inside the square brackets should
read $1+\Delta_m$.  In (137) of \cite{dan5}, $\delta\hat
S_\sigma^{(1)}$ should be dropped.  In (138) of \cite{dan5},
$\hat Y_{N_s}^{(1)}(t')$ should read $\tilde Y_{N_s}^{(1)}(t')$
and $C_s$ should read $C_s^{(b)}$.} Both $\tilde\Delta_m$,
$\tilde\Delta_{gh}$ and $U(\{\sigma_p\})-I$ are proportional to
the odd Schottky parameters. So (\ref{upsil}) and (\ref{upsilm})
can be calculated to be series over odd Schottky parameters. The
superdeterminant in (\ref{upsil}) is defined as
\begin{equation}
sdet U=\frac{\det
U_{(bb)}}{\det U_{(ff)}}\det
[I- U_{(bb)}^{-1}
U_{(bf)}U_{(ff)}^{-1} U_{(fb)}]
\label{supdet}
\end{equation}
where $U_{(bb)}$, $U_{(bf)}$, $U_{(fb)}$ and $U_{(ff)}$ are
submatrices forming the above $U$ matrix. The $b$ index labels
boson components and the $f$ index labels the fermion ones.

To present $\tilde\Delta_{gh}$ in (\ref{upsil}), we define
genus-1 Green functions $S_{\sigma,s}^{(1)}(t,t')$ as
\begin{equation}
S_{\sigma,s}^{(1)}(t,t')=Q_{\tilde\Gamma_s}(t_s)^{-2}\left
[G_b^{(1)}(z_s,z_s')
\theta_s'+\theta_sG_{(\sigma)}^{(1)}(z_s,z_s')-
\tilde\varepsilon_s'\Sigma_\sigma(z_s')\right]
Q_{\tilde\Gamma_s}(t_s')^3
\label{S1sig}
\end{equation}
where $t_s=(z_s|\theta_s)$  is defined by (\ref{gammat}), the
$Q_{\tilde\Gamma_s}$ factor is defined by (\ref{supfac}) and
$G_{(\sigma)}^{(1)}$ is $G_{(\sigma)}$ defined by
(\ref{grsig}) for genus $n=1$. Furthermore, the boson
contribution $G_b^{(1)}$ in (\ref{S1sig}) is $G_b$
taken at genus $n=1$. For an arbitrary genus-$n$,
$G_b$ is defined to be
\begin{equation}
G_b(z,z')=-\sum_\Gamma\frac{1}
{[z-g_\Gamma(z')][c_\Gamma z'+d_\Gamma]^4}
\label{gb}
\end{equation}
where the summation is performed over all the group product
of basic Schottky group elements (\ref{sch}). The
last term in (\ref{S1sig}) is defined to be limit of
$zG_{(\sigma)}^{(1)}(z_s,z_s')$ at $z\rightarrow\infty$. Owing
to this term, $S_{\sigma,s}^{(1)}(t,t')$ decreases at
$z\rightarrow\infty$ or at $z'\rightarrow\infty$.
In (\ref{upsil}), the $\Delta_{gh}(\{\sigma_r\})\}$ integral
operator is formed by the
$\{\tilde\Delta_{gh}^{(p)}(\{\sigma_r\})\}$ set of the
$\tilde\Delta_{gh}^{(p)}(\{\sigma_r\})$ integral operators, the
kernels being $\tilde\Delta_{gh}^{(p)}(\{\sigma_r\})(t,t')dt'$.
We define the kernel together with the differential
$dt'=dz'd\theta'/2\pi i$ to have deal with the objects obeying
bose statistics. Every the
$\tilde\Delta_{gh}^{(p)}(\{\sigma_r\})$ integral operator being
applied to a function of $t'$, performs integrating over $t'$
along the $C_p$ contour. The above $C_p$-contour gets around in
the positive direction both $C_{v_r}$ and $C_{u_r}$ circles
(\ref{circ}) together with the $\tilde C_p$ cut, if this cut
presents ( i.e. $l_{1p}\neq0$ ).  The $\tilde C_p$ cuts are
defined next to eq.(\ref{inf}). In the explicit form
\begin{equation}
\tilde\Delta_{gh}^{(p)}(\{\sigma_r\})(t,t')=
\int\limits_{C_p}G^0(t,t_1;\{\sigma_q\})
\frac{dz_1d\theta_1}{2\pi i}\delta
S_\sigma^{(1)}(t_1,t')
\label{Dghp}
\end{equation}
where as it was explained above, $C_p$-contour
gets around in the positive direction both circles (\ref{circ})
together with the $\tilde C_p$ cut. The Green function
$G^0(t,t_1;\{\sigma_q\})$ is defined as
\begin{equation}
G^0(t,t';\{\sigma_p\})=
G_b(z,z')\theta'+\theta G_{(\sigma)} (z,z')
\label{G0}
\end{equation}
where $G_b(z,z')$
is defined by (\ref{gb}) and  $G_{(\sigma)}(z,z')$  is given
by (\ref{grsig}).

To present the expression for
$\hat U_{N_rN_s}(\{\sigma_p\})$ in (\ref{upsil}) we define
$3/2$-tensors $\Psi_{\sigma,N_r}^{(0)}(z)$ by
\begin{equation}
\Phi_{\sigma,N_s}^{(0)}=\sum_{N_r=\mu_r,\nu_r}\hat
M_{N_s,N_r}(\{\sigma_p\}) \Psi_{\sigma,N_r}^{(0)}\quad
{\rm{where}}\quad \hat M_{N_s,N_R}(\{\sigma_p\})=
\int\limits_{C_{v_r}}
\Phi_{\sigma,N_s}^{(0)}(z)\frac{dz} {2\pi i}
\tilde Y_{\sigma,N_r}^{(1)}(z).
\label{Psis}
\end{equation}
where $\tilde Y_{\sigma,N_r}^{(1)}(z)$ is equal to
$\tilde Y_{\sigma,N_r}(z)$ defined by
eq.(\ref{ysig}) at the genus $n=1$. In this case
the $U_{N_rN_s}'(\{\sigma_p\})$ elements
of the $\hat U(\{\sigma_p\})$ matrix are given by
\begin{eqnarray}
\hat U_{N_rF_s}(\{\sigma_p\})=I-\sum_p
\int\limits_{C_p}\Psi_{\sigma,N_r}^{(0)}(z)dt
\int\limits_{C_p}\delta
S_{\sigma,p}^{(1)}(t,t_1)dt_1 \nonumber\\
\times\sum_h
\int\limits_{C_h}\tilde\Delta^{(h)}(t_1,t_2)
dt_2\int\limits_{C_{v_s}}\theta_2G_{(\sigma)} (z_2,z')dz'
\tilde Y_{F_s}^{(1)}(z')
\label{uprime}
\end{eqnarray}
where $dt=dzd\theta/2\pi i$, $G_{(\sigma)} (z,z')$ is
defined by (\ref{grsig}), $\tilde Y_{F_s}^{(1)}(t')$ is defined
by (\ref{ysig}) at the genus $n=1$ and $\delta\hat
S_\sigma^{(1)}$ is referred to those terms in (\ref{S1sig}),
which are proportional to the odd Schottky parameters.
Furthermore, $\tilde\Delta^{(h)}(t_1,t_2)dt_2$ present the
kernels of the $\tilde\Delta^{(h)}$ integral operators.  The
$\{\tilde\Delta^{(h)}\}$ set of these operators forms the
$\Delta$  operator that can be given as
\begin{equation}
\Delta=[I+\tilde\Delta_{gh}(\{\sigma_p\})]^{-1}
\label{tildel}
\end{equation}
where the $\tilde\Delta_{gh}(\{\sigma_p\})$ operator is the same
as in (\ref{upsil}). Eqs. (\ref{upsil} and (\ref{uprime}
correspond to eqs. (137) and (138) of \cite{dan5} with $\hat U$
to be $U'$ of \cite{dan5}. But in (\ref{upsil} and (\ref{uprime}
both $\Upsilon_{gh}^{(n)}(\{q_{N_s}\},L)$ and $\hat U$ is given
in terms of $\tilde\Delta_{gh}$ instead of $\Delta_{gh}$ defined
in \cite{dan5}. This leads to more compact, formulas, especially
for the $U(\{\sigma_p\})$ matrix presented below. The proof of
(\ref{upsil}) and (\ref{uprime}) is achieved by an expansion in
powers of $\Delta_{gh}$ of (137) and of (138) in \cite{dan5}.In
this case sum of integrations over $t'$ of every particular
$\delta S_{\sigma,p}^{(1)}(t,t')$ along the $C_r$ contours
$(r\neq p)$ is reduced to the integral along the $C_p$ contour.
As the result, eqs.  (\ref{upsil} and (\ref{uprime} arise.

To present the $U(\{\sigma_p\})$ matrix in (\ref{upsil}) we
define $3/2$-supertensors $\Psi_{\sigma,N_r}^{(1)}(z)$ on the
genus-$1$ supermanifolds by
\begin{equation}
S_{\sigma,s}^{(1)}(t_s^b,t')=Q_{\Gamma_{b,s}}^{-2}(t)
\left(S_{\sigma,s}^{(1)}(t,t')+\sum_{N_s}
\hat Y_{\sigma,N_s}^{(1)}(t)
\Psi_{\sigma,N_s}^{(1)}(t')\right)
\label{psi1}
\end{equation}
where $N_s=(k_s,u_s,v_s,\mu_s,\nu_s)$ and the $t\rightarrow
t_s^b$ transformation is defined in (\ref{stens}).  For
$N_s=(k_s,u_s,v_s)$, the $\hat Y_{b,N_s}^{(1)}(t)$ polynomials
in (\ref{psi1})  are equal to $P_{R_s}(z_s)
Q_{\tilde\Gamma_s}^{-2}(t_s)$ where $P_{R_s}$ are defined by
(\ref{poly}).  For $N_s=(\mu_s,\nu_s)$, the above $\hat
Y_{\sigma,N_s}(t)$ polynomials are equal to $\tilde
Y_{p,N_r}(t_s)Q_{\tilde\Gamma_s}^{-2}(t_s)$, $\tilde
Y_{\sigma,N_r}(t)$ being defined by (\ref{ysig}) at the genus
$n=1$. And $Q_{\tilde\Gamma_s}(t_s)$  is defined for $t_s$
transformation (\ref{gammat}) by (\ref{supfac}). Furthermore, we
define $\Psi_{\sigma,N_s}(t)$ to be an extension of
$\Psi_{\sigma,N_s}^{(0)}(z)$ to nonzero odd moduli as follows
\begin{equation}
\Psi_{\sigma,N_s}(t)=\int\limits_{C_s}
\Psi_{\sigma,N_s}^{(1)}(t')\frac{dz'd\theta'}{2\pi
i}S_\sigma(t',t)
\label{psi}
\end{equation}
where the Green function $S_\sigma(t,t')$ is given by
\begin{eqnarray}
S_\sigma(t,t')=\tilde G(t,t';\{\sigma_p\})+\sum_h
\int\limits_{C_h}\tilde\Delta^{(h)}(t',t_1)
dt_1\int\limits_{C_{v_s}}G_{(\sigma)} (z_1,z_2)dz_2
\tilde Y_{F_s}^{(1)}(z_2)\hat U_{F_sN_r}^{-1}\nonumber\\
\times\left[\sum_p\int\limits_{C_p}\Psi_{\sigma,N_r}^{(0)}(z)dt_3
\int\limits_{C_q}\delta
S_{\sigma,p}^{(1)}(t_3,t_4)dt_4\tilde
G(t_4,t';\{\sigma_p\})-\Psi_{\sigma,N_r}^{(0)}(t')\right]
\label{sss}
\end{eqnarray}
where $\Psi_{\sigma,N_r}^{(0)}(t')$ is given by (\ref{Psis}),
$\hat U_{F_sN_r}$ is given by (\ref{uprime}) and $\tilde
G(t_4,t';\{\sigma_p\})$ is defined as
\begin{equation}
\tilde G(t,t';\{\sigma_p\})=
\sum_h\int\limits_{C_h}\tilde
\Delta^{(h)}(t,t_1) dt_1G^0(t_1,t';\{\sigma_p\})
\label{ggg}
\end{equation}
In this case the desired $\tilde U_{N_sN_r}(\{\sigma_p\})$ can
be given as
\begin{equation}
\tilde U_{N_sN_r}(\{\sigma_p\})=
\int\limits_{C_{v_r}}
\Psi_{\sigma,N_s}(t)\frac{d\theta dz} {2\pi i}
P_{N_r}(t)+\int\limits_{\tilde C_r}
\Psi_{\sigma,N_s}(t)\frac{d\theta dz} {2\pi i}
P_{N_r}^{(a)}(t)
\label{uuu}
\end{equation}
where both $P_{N_r}(t)$ and $P_{N_r}^{(a)}(t)$ are defined by
(\ref{poly}) and (\ref{pola}) and $\Psi_{\sigma,N_s}(t)$ is
given by (\ref{psi}).  It is useful to note that both
$S_\sigma(t,'t)$ and $\Psi_{\sigma,N_s}(t)$ of this paper are
the same as in \cite{dan5}. The proof of (\ref{psi}) is quite
similar to that of eq.(54) in \cite{dan5}. Eq.(\ref{sss}) is the
explicit form of the solution of eq.(84) of \cite{dan5} given in
terms of the $G^0(t,t';\{\sigma_p\})$ Green function (\ref{G0}).
Eqs. (\ref{upsil}), (\ref{uprime}) and (\ref{uuu}) give the
desired $\Upsilon_{gh}^{(n)}(\{q_{N_s}\},L)$ factor in
(\ref{b3}).

The $\Upsilon_m^{(n)}(\{q_{N_s}\},L)$ in(\ref{b3}) is determined
by (\ref{upsilm}) in terms of the $\tilde\Delta_m$ integral
operator. To present the above operator we consider the
holomorphic Green functions $R_L(t,t')$ of the scalar superfields
and the kindred Green functions $K_L(t,t')$ defined to be
\begin{equation}
K_L(t,t')=D(t')R_L(t,t')
\label{klrl}
\end{equation}
where $D(t)$ is the spinor derivative (\ref{supder}). The
periods of $R_L(t,t')$ are $J_s(t;L)$ and the periods of
$J_s(t;L)$ form the $2\pi i\omega(\{q_N\},L)$ matrix,
$\omega(\{q_N\},L)$ being the period matrix in (\ref{hol}).
It can be shown \cite{dan5} that $K_L(t,t')$ obeys the integral
equation  with the kernel to be none other that the kernel of
the desired integral operator $\tilde\Delta_m$ in
(\ref{upsilm}).

To give in the explicit form the kernel of $\tilde\Delta_m$ one
can note that for zero odd modular parameters, $R_L(t,t')$ is
reduced to $R_{(0)}(t,t';L)$ as
\begin{equation}
R_{(0)}(t,t';L)=R_b(z,z')-\theta\theta'R_f(z,z';L)
\label{rr0}
\end{equation}
where $R_b(z,z')$ is the boson Green function and $R_f(z,z';L)$
is the fermion Green one. The $R_b(z,z')$ Green function is
given  by \cite{fried}
\begin{equation}
R_b(z,z')=\sum_\Gamma\ln\left(\frac{[z-g_\Gamma(z')][-c_\Gamma
z^{(o)}+ a_\Gamma]}{[-c_\Gamma
z+a_\Gamma][z^{(o)}-g_\Gamma(z^{(1)})]}\right)
\label{rrr0}
\end{equation}
$z^{(o)}$ and  $z^{(1)}$ being arbitrary constants.  The fermion
Green function $R_f(z,z';L)$ in (\ref{rr0}) can be given as
\begin{equation}
R_f(z,z';L)=\exp\left\{\frac{1}{2}[R_b(z,z)+R_b(z',z')]-
R_b(z,z')\right\}
\frac{\Theta[l_1,l_2](J|\omega^{(r)})}
{\Theta[l_1,l_2](0|\omega^{(r)})}
\label{ff0}
\end{equation}
where Green function $R_b(z,z)$ for $z'=z$ is defined to be
the limit of  $R_b(z,z')-\ln(z-z')$  at $z\rightarrow z'$.
Furthermore, $\Theta$ is the theta function and the symbol $J$
denotes the set of functions
$(J_{(0)s}(z)-J_{(0)s}(z'))/2\pi i$, $J_{(0)s}(z)$ being
periods of $R_b(z,z')$. We define also for arbitrary odd moduli
the genus-1 Green functions $R_s^{(1)}(t,t')$  as
\begin{equation}
R_s^{(1)}(t,t') =R_{(0)s}^{(1)}(t_s,t_s')+
\tilde\varepsilon_s'\theta_s'\Xi_s(\infty,z_s')-
\theta_s\tilde\varepsilon_s'\Xi_s(z_s,\infty)
\quad{\rm for}\quad
s=1,2,...n
\label{rs1}
\end{equation}
where both  $t_s=(z_s|\theta_s)$,  $t_s'=(z_s'|\theta_s')$ and
$\tilde{\varepsilon}_s'$ are defined by (\ref{gammat}) and
$R_{(0)s}^{(1)}(t,t')$ is $R_s^{(1)}(t,t')$ at zero odd moduli.
Furthermore, $\Xi_s(z,z')$ is
\begin{equation}
\Xi_s(z,z')=(z-z')R_{(f)s}^{(1)}(z,z';l_{1s},l_{2s})
\label{xis}
\end{equation}
Two the last terms in (\ref{xis})  provide decreasing
$K_s^{(1)}(t,t')$ at $z\rightarrow\infty$ or
$z'\rightarrow\infty$ where $K_s^{(1)}(t,t')$ is defined by
eq.(\ref{klrl}) for $R= R_s^{(1)}$. Being twisted  under
$(A_s,B_s)$-circles, $R_s^{(1)}(t,t')$ is changed by
$(\Gamma_{a,s},\Gamma_{b,s})$-mappings (\ref{super}).
To calculate $R_{(0)s}^{(1)}(t,t')$ in (\ref{rs1}) for even
genus-1 spin structures we use (\ref{rrr0}) and(\ref{ff0}) at
$n=1$. The genus-1 spin structure being odd, we defined
$R_{(f)s}^{(1)}(z,z')$ as
\begin{equation}
R_{(f)s}^{(1)}(z,z')=
\frac{\partial_z\{\Theta[1/2,1/2](J_{(1)}|\omega_s^{(1)})\}}
{\Theta[1/2,1/2](J_{(1)}|\omega_s^{(1)})}
\sqrt{\frac{\partial_{z'}J_{(0)s}^{(1)}(z')}
{\partial_zJ_{(0)s}^{(1)}(z)}}
\label{ffod}
\end{equation}
where $\Theta$ is
the genus-1 theta
function. Furthermore,
$J_{(1)}=(J_{(0)s}^{(1)}(z)-J_{(0)s}^{(1)}(z'))/2\pi i$ and
$J_{(0)s}^{(1)}$ is the period of $R_{(b)s}^{(1)}(z,z')$,
the period of $J_{(0)s}^{(1)}$ being $2\pi i\omega_s^{(1)}$.
In this case, for every $s$, the  Green function
$K_s^{(1)}(t,t')=D(t')R_s^{(1)}(t,t')$  is changed under
$\Gamma_{b,s}$ transformation as
\begin{eqnarray}
K_s^{(1)}(t,t_s^b(t'))=\left[K_s^{(1)}(t,t')+
\varphi_s(t)f_s(t')\right]Q_{\Gamma_{b ,s}}(t')\nonumber\\
K_s^{(1)}(t_s^b(t),t')=K_s^{(1)}(t,t')+
2\pi i\eta_s^{(1)}(t')-
\varphi_s(t)f_s(t')
\label{k1ch}
\end{eqnarray}
where  $f_s(t')=D(t')\varphi_s(t')$. The above $\varphi_s(t')$
disappears, if $(l_{1s},l_{2s})$-characteristics correspond to
an even genus-1 spin structure. In this case the
desired integral equation for $K_L(t,t')$ has the following form
\cite{dan5}
$$K_L(t,t')=K_{(0)}(t,t';L)-
\sum_{r=1}^{n}\int\limits_{C_r}K_{(0)}(t,t_1;L)
dt_1\delta K_r^{(1)}(t_1,t_2)dt_2K_L(t_2,t')-
\int\limits_{C_r}
K_{(0)}(t,t_1;L) $$
\begin{equation}
\times\delta\varphi_r(t_1)dt_1\int
\limits_{\hat C_{v_r}}f_r(t_2)dt_2
K(t_2,t')+
\int\limits_{ C_{v_r}}K_{(0)}(t,t_1;L)
\varphi_{(0)r}(t_1)dt_1
\int\limits_{C_r}\delta f_r(t_2)dt_2K(t_2,t')
\label{kleq}
\end{equation}
where $dt=d\theta dz/2\pi i$, etc.  Furthermore,
$K_{(0)}(t,t_1;L)$ denotes $K_L(t,t_1)$ taken at all the odd
Schottky parameters to be zero. And $\varphi_{(0)r}(t_1)$ is
equal to $\varphi_r(t_1)$ at $\mu_r=\nu_r=0$. Each of $\delta
K_r^{(1)}, \delta\varphi_r$ and $\delta f_r$ is defined to be
the difference between the corresponding value and that
calculated for zero values of odd Schottky parameters
$(\mu_r,\nu_r)$.  As example, $\delta
K_r^{(1)}(t_1,t_2)=K_r^{(1)}(t_1,t_2)- K_{(o)r}^{(1)}(t_1,t_2)$
where $K_{(o)r}^{(1)}(t_1,t_2)$ is the $K_r^{(1)}(t_1,t_2)$
function taken at $\mu_r=\nu_r=0$. The $C_r$ contours are
defined as in (\ref{Dghp}). On the $z_s$ complex plane
(\ref{gammat}) the $\hat C_{v_s}$ contour is none other than the
$C_{v_s}$ circle (\ref{circ}). Only the odd genus-1 spin
structures contribute in two last terms on the right side of
(\ref{kleq}). The term $K_{(0)}(t,t';L)=D(t')R_{(0)}(t,t';L)$
outside the integral on the right side of (\ref{kleq}) is
calculated in the terms of both $R_b(z,z')$ and $R_f(z,z';L)$,
as it has been explained above. Since the  kernel of
(\ref{kleq}) is proportional to odd parameters, solution of
(\ref{kleq}) can be obtained by the iteration procedure, every
posterior iteration being, at least, one more power in odd
parameters than a previous one.  Therefore, $K_L(t,t')$ appears
to be a series containing a finite number of terms.  After
$K_L(t,t')$ being determined, the $R_L(t,t')$ Green function  is
calculated without essential difficulties.

The kernel of integral operator $\tilde\Delta_m$ in
(\ref{upsilm}) is just the kernel of (\ref{kleq}). The right
side of (\ref{upsilm}) is calculated by an expansion in
powers of $\tilde\Delta_m$. Eq. (\ref{upsilm} is more convenient
for the calculation than eq.(134) of \cite{dan5} where
$\Upsilon_m^{(n)}(\{q_{N_s}\},L)$ is given in terms
of $\Delta_m$ defined by (135) in \cite{dan5}.  To prove
identity of (\ref{upsilm}) with (134) of \cite{dan5} one can
verify that
\begin{equation}
trace\ln\left[1+\tilde\Delta_m\right]-
trace\ln\left[1+\Delta_m\right]=0\,.
\label{dmtdm}
\end{equation}
The proof of (\ref{dmtdm}) is achieved
by an expansion in powers of both $\Delta_m$ and $\Delta_m$,
eqs. (50) and (51) of \cite{dan5} being used. It is also
employed that sum of integrations over $t_1$ of every
particular $\delta K_s^{(1)}(t,t_1)$ along the $C_r$ contours
$(r\neq s)$ is reduced to the integral along $C_s$-contour.

The $J_p$ periods of
$R_L(t,t')$ in (\ref{omeq}) are calculated as \cite{dan5}
\begin{equation}
J_p(t;L)=\int\limits_{C_p}K(t,t')J_p^{(1)}(t')\frac{d\theta'dz'}
{2\pi i}
\label{js}
\end{equation}
where $J_r^{(1)}(t')$ is the period of the genus-1 Green
function $R_r^{(1)}(t,t')$.  In (\ref{omeq}) and (\ref{js}) the
integration contour $C_r$ is defined as in (\ref{Dghp}).
The $\omega_{rp}$ matrix elements of the
period matrix $\omega(\{q_N\},L)$ in the measure (\ref{hol})
can be calculated as \cite{dan5}
\begin{equation}
2\pi i\omega_{rp}=k_r\delta_{rp}+
\int\limits_{C_r}D(t)J_p(t;L)J_r^{(1)}(t)\frac{d\theta dz}
{2\pi i}
\label{omeq}
\end{equation}
where $k_r$ is the Schottky multiplier. The right side of
(\ref{omeq}) can be proved to be symmetrical in respect to
interchanging $r$ and $p$.

For all the $l_{1s}$ theta characteristics to be zero
( that is the Neveu-Schwarz sector ) eq.(\ref{b3}) is reduced to
(\ref{znsch}) and (\ref{omeq}) is reduced to (\ref{b1}).

\newpage
\begin{center}
{\bf Figure} in {\bf G.S. Danilov}'s paper {\bf"Unimodular
transformations of the supermanifolds and the calculation of the
multi-loop amplitudes  in the superstring theory"}
\end{center}

\begin{figure}
\setlength{\unitlength}{1mm}
\begin{picture}(160,60)
\multiput(8,50)(58,0){3}{\circle{16}}
\multiput(40,50)(58,0){2}{\circle{16}}
\multiput(41,25)(58,0){2}{\circle{10}}
\multiput(12,25)(58,0){3}{\circle{10}}

\multiput(15,25)(58,0){3}{\circle*{1}}
\multiput(38,25)(58,0){3}{\circle*{1}}
\multiput(11,50)(58,0){3}{\circle*{1}}
\multiput(37,50)(58,0){3}{\circle*{1}}

\multiput(11,50)(116,0){2}{\line(1,0){26}}
\multiput(15,25)(116,0){2}{\line(1,0){23}}
\put(95,50){\line(-1,0){5}}
\put(69,50){\line(1,0){10}}
\put(85,50){\line(0,-1){23}}
\put(73,25){\line(1,0){10}}
\put(82,27){\oval(6,4)[br]}
\put(96,27){\oval(6,4)[bl]}
\put(97,52){\oval(18,12)[t]}
\put(96,50){\oval(22,18)[t]}
\put(95,53){\oval(26,18)[t]}

\put(97,52){\oval(18,20)[br]}
\put(96,50){\oval(22,18)[br]}
\put(96,39){\oval(6,4)[tl]}
\put(93,27){\line(0,1){13}}

\put(79,53){\oval(6,6)[br]}
\put(91,52){\oval(6,4)[bl]}

\put(97,29){\oval(18,26)[l]}
\put(97,29){\oval(22,26)[br]}
\put(108,29){\line(0,1){24}}

\put(15,25){\makebox(0,0)[tr]{$v_2$}}
\put(73,25){\makebox(0,0)[tr]{$v_2$}}
\put(131,25){\makebox(0,0)[tr]{$v_2$}}

\put(36,26){ {$u_2$}}
\put(94,26){ {$u_2$}}
\put(152,26){ {$u_2$}}

\put(11,50){\makebox(0,0)[tr]{$v_1$}}
\put(69,50){\makebox(0,0)[tr]{$v_1$}}
\put(127,50){\makebox(0,0)[tr]{$v_1$}}

\put(35,51){ {$u_1$}}
\put(93,51){ {$u_1$}}
\put(151,51){ {$u_1$}}

\put(24,0){\makebox(0,0){(a)}}
\put(82,0){\makebox(0,0){(b)}}
\put(140,0){\makebox(0,0){(c)}}
\thicklines
\put(156,50){\circle{16}}
\put(157,25){\circle{10}}
\put(152,27){\oval(6,4)[bl]}
\put(149,50){\line(0,-1){23}}
\put(151,38){\makebox(0,0){r}}
\put(147,38){\makebox(0,0){l}}

\end{picture}
\vspace{0.5cm}
\caption{The going of $C_{u_2}$ circle round the
$C_{u_1}$ one: the initial position (a), the final position (b),
the cuts are reduced to be closed together (c).}
\end{figure}

\newpage
\begin{thebibliography}{99}
\bibitem{rnshw}
P. Ramond,   Phys.Rev. D3 (1971) 2415.\\
A. Neveu  and J.H. Schwarz,   Nucl.Phys. B31 (1971) 86.
\bibitem{gosch}
F. Gliozzi, D. Olive and J. Scherk,   Phys.Lett. 65B (1976) 282.
\bibitem{bshw}
M.A. Baranov and A.S.
Schwarz, Pis'ma ZhETF 42 (1985) 340 [JETP Lett. 49 (1986) 419];
D. Friedan, Proc. Santa Barbara Workshop on Unified
String theories, eds. D. Gross and M. Green ( World Scientific,
Singapore, 1986).
\bibitem{swit}
N. Seiberg and E. Witten, Nucl.Phys. B276 (1986) 272.
\bibitem{ver}
E. Verlinde and H. Verlinde, Phys. Lett.  B 192 (1987) 95.
\bibitem{martnp}
E.  Martinec, Nucl.  Phys. B 281 (1986) 157.
\bibitem{as}
J. Atick and A. Sen, Nucl.Phys. B. 296 (1988) 157;\\ J. Atick, J.
Rabin and A.  Sen, Nucl. Phys. B 299 (1988) 279.
\bibitem{momor}
G. Moore and A. Morozov, Nucl.  Phys. B 306 (1988) 387; \\
A. Yu.  Morozov, Teor.  Mat.  Fiz. 81 (1989) 24.
\bibitem{vec8}
P. Di Vecchia, K. Hornfeck, M. Frau, A. Lerda and S. Sciuto,
Phys. Lett.  B 211 (1988) 301.
\bibitem{pst}
J.L. Petersen, J.R. Sidenius and A.K. Tollst{\'e}n,
Phys Lett. B 213 (1988) 30;
Nucl.  Phys.  B 317 (1989) 109.
\bibitem{ntw}
B.E.W. Nilsson, A.K. Tollst{\'e}n and A. W{\"a}tterstam,
Phys Lett. B 222 (1989) 399.
\bibitem{dan1}
G.S. Danilov, Phys.  Lett. B 257 (1991) 285;
\bibitem{dan0}
G.S. Danilov, Sov. J. Nucl.  Phys. 52 (1990) 727
[ Jadernaya Fizika 52 (1990) 1143 ].
\bibitem{dan5}
G.S. Danilov, Phys. Rev. D 51(1995)4359.
\bibitem{martpl}
E.  Martinec,  Phys.Lett. B171 (1986) 189.
\bibitem{mandel}
S. Mandelstam, Phys.Lett. B 277 ( 1992 ) 82.
\bibitem{berk}
N. Berkovits, Nucl. Phys. B408 (1993) 43.
\bibitem{crrab}
L. Crane and J.M. Rabin , Commun. Math. Phys. 113 (1988) 601;\\
J.D. Cohn, Nucl. Phys. B306 (1988) 239.
\bibitem{leites}
D.A. Leites, Usp. Mat. Nauk, 35 (1980) 1.
\bibitem{rschv}
A.A. Rosly, A.S. Schwarz and A.A. Voronov, Commun. Math. Phys.
119 (1986) 129.
\bibitem{cqg}
G.S. Danilov, Class. Quantum Grav. 11 (1994) 2155.
\bibitem{lovel}
C. Lovelace, Phys. Lett.
B 32 (1970) 703; V.  Alessandrini, Nuovo Cim. 2A (1971) 321; V.
Alessandrini and D. Amati, Nuovo Cim. 2A (1971) 793.
\bibitem{fried}
D. Friedan, E. Martinec and S. Shenker,
Nucl. Phys. B 271 (1986) 93.
\bibitem{hodkin}
L. Hodkin, J. Physics and Gravity, 6 (1989) 333.
\bibitem{dan3}
G.S. Danilov, JETP Lett. 58 (1993) 796 [ Pis'ma JhETF 58 (1993)
790. ]
\bibitem{dan4}
G.S. Danilov, Physics of Atomic Nuclei 57 (1994) 2183
[ Jadernaya Fizika 57 (1994) 2272 ].
\bibitem{vec7}
P. Di Vecchia, M. Frau, A. Lerda and S.
Sciuto, Phys.  Lett.  B 199 (1987) 49.
\bibitem{dan9}
G.S. Danilov,  Sov. J. Nucl. Phys. 49 (1989) 1106
[ Jadernaya Fizika 49 (1989) 1787].
\bibitem{pol}
A.M. Polyakov,
Phys. Lett. B 103 (1981) 210; Phys. Lett. B 103 (1981) 207.
\bibitem{davis}
S. Davis, DAMTP-R-94-27 (July, 1994), hep-th/ 9505231.
\bibitem{belkniz}
A.A. Belavin and V.G. Knizhnik, Phys. Lett. B 168 (1986) 201.


\end{thebibliography}
\end{document}